\documentclass[12pt,preprint]{aastex}

\usepackage{graphicx}

\shorttitle{Ion injection decoded}
\shortauthors{Sundberg et al.}

\begin{document}


\title{Ion Acceleration at the Quasi-Parallel Bow Shock: Decoding the Signature of Injection}


\author{Torbj\"{o}rn Sundberg, Christopher T. Haynes, David Burgess}
\affil{School of Physics and Astronomy, Queen Mary University of London, London, E1 4NS, United Kingdom}

\author{Christian X. Mazelle}
\affil{University Paul Sabatier Toulouse III, Toulouse Cedex 09, France}




\begin{abstract}
Collisionless shocks are efficient particle accelerators. At Earth, ions with energies exceeding 100 keV are seen upstream of the bow shock when the magnetic geometry is quasi-parallel, and large-scale supernova remnant shocks can accelerate ions into cosmic rays energies.  This energization is attributed to diffusive shock acceleration, however, for this process to become active the ions must first be sufficiently energized. How and where this initial acceleration takes place has been one of the key unresolved issues in shock acceleration theory. Using Cluster spacecraft observations, we study the signatures of ion reflection events in the turbulent transition layer upstream of the terrestrial bow shock, and with the support of a hybrid simulation of the shock, we show that these reflection signatures are characteristic of the first step in the ion injection process. These reflection events develop in particular in the region where the trailing edge of large-amplitude upstream waves intercept the local shock ramp and the upstream magnetic field changes from quasi-perpendicular to quasi-parallel. The dispersed ion velocity signature observed can be attributed to a rapid succession of ion reflections at this wave boundary. After the ions' initial interaction with the shock, they flow upstream along the quasi-parallel magnetic field. Each subsequent wave front in the upstream region will sweep the ions back toward the shock, where they gain energy with each transition between the upstream and the shock wave frames. Within three to five gyroperiods, some ions have gained enough parallel velocity to escape upstream, thus completing the injection process.

\end{abstract}

\section{Introduction}

Collisionless shocks are found in a range of different astrophysical environments, where they provide the dissipation necessary to connect a supersonic flow upstream of the shock to a downstream subsonic flow on a scale much shorter than the collisional mean free path of the plasma. In our solar system, shocks can be found upstream of all the planets, ahead of coronal mass ejections, around solar wind stream interactions, and at the outer boundary of the Heliosphere. They are also believed to exist in a wide variety of astrophysical contexts, such as, for example, around supernova remnants. 

At a high Mach number shock, the flow energy in the upstream plasma is mainly converted to thermal energy of the particles, leading to a slowed, thermalized plasma downstream of the shock. The required dissipation primarily takes place through interaction between the particles and the electric and magnetic fields at the shock interface, together with wave-particle interactions in the wave field driven by instabilities at the shock. In addition, due to the lack of collisions, a small fraction of particles can reach high energies via the shock fields and coupling to the flow via wave-particle interactions. The population of energetic particles associated with a collisionless shock can have an important effect on its overall behavior.

The two main controlling factors of the shock structure and dynamics are the Alfv\'{e}nic Mach number, expressed as the ratio between the upstream flow speed normal to the shock surface and the upstream Alfv\'{e}n velocity,  $M_A=V_u/V_A =  V_u/(B_u/\sqrt{\mu_0 \rho_u})$, and the inclination of the upstream magnetic field to the shock normal, $\theta_{Bn}$. Here, $V_u$ is the upstream flow velocity in the direction of the shock normal, $V_A$ is the Alfv\'{e}n velocity, $B_u$  is the upstream magnetic field, $\mu_0$ is the permeability of free space, and $\rho_u$ is the upstream mass density. Other factors such as the sonic Mach number and the shock curvature may also have an influence on the shock dynamics, but typically to a lesser extent. The orientation of the upstream magnetic field separates shocks into two qualitatively different configurations: quasi-parallel, $\theta_{Bn}<45^\circ$, and quasi-perpendicular, $\theta_{Bn}\ge45^\circ$. At both quasi-perpendicular and quasi-parallel shocks specular (or near-specular) reflection of some fraction of the incident ions is found be to crucial for understanding ion heating and acceleration \citep{gosling1985}. In a quasi-perpendicular geometry, ions that have been specularly reflected from the incident flow at the shock front are bound to return to the shock within one gyro-orbit, whereas at a quasi-parallel shock, such reflected ions can flow back along the magnetic field from the shock surface and reach far into the upstream region, where they populate the ion foreshock, generate turbulence, and initiate wave growth. Consequently, the region upstream of a quasi-parallel shock is self-seeded with upstream waves that generate a dynamic shock layer, often including features such as whistlers, ultra-low frequency (ULF) waves, shocklets and other large-amplitude magnetic structures. This is also the region which is most intimately linked to the generation of the high-energy upstream ions, and in particular the extraction of thermal particles into the population of energetic ions. 

Ion acceleration at the shock can be caused by shock drift acceleration \citep[e.g., ][]{armstrong1985, burgess1987, decker1988}, diffusive shock acceleration \citep[e.g., ][]{axford1977,krymskii1977,bell1978-I, bell1978-II, blandford1978}, and wave-particle interaction \citep[e.g., ][]{sugiyama1999, mazelle2003, kuramitsu2005}. In the solar system, shock drift acceleration is usually considered to be the main mechanisms operating at quasi-perpendicular shocks, in the case where the surrounding turbulence has low amplitude. In this case the ions can gyrate across the shock transition layer a few times, each time gaining additional energy in the velocity component perpendicular to the magnetic field, until they eventually escape from the shock. This results in a heated and anisotropic ion distribution downstream of the shock, and little energization of upstream ions. The generation of high-energy upstream ions is instead considered a result of diffusive shock acceleration. This theory relies on the diffusion of energetic ions in both the upstream and downstream regions so that ions are continuously returned to the shock, leading to higher and higher ion energies after each shock crossing. For more details, see reviews by \citet{drury1983}, \citet{jones1991}. At supernova remnant shocks, which are fast shocks with very large spatial scales, such ions are believed to be accelerated up to GeV energies and above. As both the energy levels and the predicted energy spectrum agrees well with cosmic ray observations \citep[e.g., ][]{berezhko2005, berezhko2007, ptuskin2010, schure2012, blasi2013}, diffusive shock acceleration at supernova remnant shocks is considered as the major source of galactic cosmic rays. Computer simulations give additional support for these types of energy distributions, with both thermal and non-thermal ion distributions being generated at the shock front \citep[e.g., ][]{giacalone1992, giacalone2004, gargate2012}.

In order for diffusive shock acceleration to become operative, ions need to undergo an initial acceleration processes to reach the energies sufficient to diffuse across the shock transition. The extraction of thermal ions to these energies is known as the injection problem. Which particular processes generate this initial energization is still undetermined, and it is the main missing link in our understanding of the ion acceleration processes at collisionless shocks. One prominent path is ion reflection at the shock front. Electric and magnetic forces in the shock transition layer are believed to cause a near-specular reflection of a fraction of the upstream flow, resulting in cold reflected ion beams, bunched in gyrophase \citep{leroy1981, leroy1982, gosling1982, gosling1985}.  In perfect specular reflection, the ion velocity component normal to the shock is reversed, while that transverse to the shock normal is preserved. In a field-aligned coordinate system with an upstream flow velocity $\bar{v}_u=[v_{u\parallel},0,v_{u\perp}]$, the reflected beam can be expressed as combination of a gyro-center motion and a gyration, $\bar{v}_r = \bar{v}_{gc} + \bar{v}_g$ \citep{gosling1982, gosling1985}, with
\begin{equation}\
	\bar{v}_{gc} =[v_{u\parallel}-2\cos(\theta_{Bn})\cos(\theta_{Vn})|v_u|, 0, v_{u\perp}], 
\end{equation}
and
\begin{equation}
	\bar{v}_{g} = -2 \cos(\theta_{Vn})|\bar{v}_u|(\bar{n}-\cos(\theta_{Bn})\bar{b}),
\end{equation}
where $\bar{n}$ is the shock normal and $\bar{b}$ is the magnetic field vector.	The ion gyrophase changes with distance from the shock, gyrating about the projection of the center of mass to the $v_{r\parallel}$ plane in a left-handed sense. The reflection of incoming ions leads to an immediate energization in the upstream reference frame, and it can provide a significant fraction of the energy needed for ion injection. 
%
%
This description of the reflection process is given in the local shock frame, i.e. the frame in which the shock is, on average, at rest. However, since ion reflection events are transient, it might be that the appropriate frame is an instantaneous shock frame which takes into account shock motion; simulations indicate that the shock frequently stalls in association with shock reformation leading to average slower shock speeds during reflection events. In these circumstances, a downstream reference frame may be more appropriate, as argued by \citet{caprioli2015}. Observationally, there are many difficulties with measuring either average or instantaneous shock speed, since gross motions of the bow shock are combined with variations on both short time and length scales.

The observational evidence for such reflected ion beams has primarily been reported from close vicinity to either the shock ramp or shock-like features. After the initial period following reflection, these are expected to spread in velocity space, both in energy and phase. This velocity space dispersion should take place on time scales shorter than the ion gyroperiod in the upstream magnetic field, evidenced by the clear dominance of ion beams at a gyrophase close to the initial value predicted by specular reflection \citep{onsager1990}. Such coherent reflection has been reported from both quasi-parallel and quasi-perpendicular sections of the terrestrial bow shock, but following the reasoning above, only quasi-parallel beams are expected to escape upstream. These observations provide general support for the importance of specular reflection for the energy dissipation and ion injection at quasi-parallel shocks, but the final fate of these ion beams is currently unknown. 

Further support for the necessity of ion reflection for injection comes from simulations, which typically show that all energetic particles originate from reflection at the shock and that they gain their energy in the immediate vicinity of the shock front \citep{scholer1990-I, scholer1990-II, kucharek1991, giacalone1992, guo2013}, rather than through leakage from the downstream medium. Additional processes that can provide ion energization are wave-shock interaction models, such as that suggested by \citet{sugiyama1999}, or gyro-resonant surfing \citep{kuramitsu2005-I, kuramitsu2005-II}. These results present strong evidence that specular reflection provides the first important step toward ion injection. 

Simulations by \citet{caprioli2014-I} have highlighted a strong requirement of quasi-parallel shock geometries for ion injection; they report that the energetic ions represent 10--20\% of the kinetic energy of the shock, for shocks with Mach numbers 10 and above. This ratio is practically independent of $\theta_{Bn}$ for quasi-parallel configurations, whereas ion injection at quasi-perpendicular shocks is  negligible. These results are also in agreement with earlier parameter studies by \citet{giacalone1997} and \citet{gargate2012}. 
%
A minimal model of the ion--shock interaction was recently proposed by \citet{caprioli2015}. By tracing ions interacting with a periodically reforming shock potential barrier that varies between a high and a low state, they show how reflection off a steep shock potential is a necessary but not sufficient requirement for ion injection. In successive shock interactions, a fraction of the initially reflected ions penetrate into the downstream region, forming a population of supra-thermal ions. Sufficient energy gain for an ion to escape into the upstream region is typically acquired after multiple shock interactions. This model can successfully explain the physical processes involved in the ion injection and the ion energy spectrum at the shock.

The present study aims to further these results by providing a link between the spacecraft observations and computer simulations. Using in-situ data at the terrestrial bow shock from the four-spacecraft Cluster mission, we study the velocity space distribution of reflected ion beams, providing a full three-dimensional velocity space view of the ion distribution, compared to the two-dimensional projection utilized in previous studies \citep{gosling1985, onsager1990}. The results of this analysis show that rather than the standard picture of cold coherent beams, these reflected ion beams typically show a spread in both energy and gyrophase already near the initial reflection point. With the aid of a hybrid simulation of an oblique shock, we will show that these signatures are consistent with ion injection events. By tracing the time history of the injected ions, we can also present a description of the injection process at quasi-parallel shocks consistent with the observations.

\section{Observations}

At Earth, the quasi-parallel bow shock is characterized by a turbulent shock transition layer and a strong upstream ULF wave field. This region is typically populated by both high-frequency whistler waves \citep{fairfield1974, greenstadt1995, burgess1997} and large-amplitude low-frequency pulsations with time periods of $\sim$30 s and a fractional magnetic field increase $\Delta B/B\ge1$.  Wave bursts at $\sim$3 s periods are also detected under certain conditions \citep{le1992}. The long-period pulsations are generated by backstreaming ions in the foreshock, their wave phase velocities are directed upstream in the plasma frame, but they are swept back toward the shock with the solar wind flow \citep[e.g.,][]{eastwood2005}. These pulsations can grow into large-amplitude structures in the area immediately upstream of the shock (sometimes referred to as short large-amplitude magnetic structures, or SLAMS), which can trigger a reformation of the shock layer, contributing to the turbulent appearance of the shock transition region \citep{schwartz1991-I,schwartz1991-II, schwartz1992}. This upstream ULF wave field also leads to variations in the instantaneous $\theta_{Bn}$ of the shock, which locally changes the dynamics of the shock, leading to short-scale variations between parallel and perpendicular conditions.

Figure 1 shows an interval of magnetic field observations by the Cluster 1 (C1) spacecraft at a crossing of the terrestrial shock on 30 March 2006. The magnetic field trace shown is typical of that of a quasi-parallel shock. The most striking feature is a repeated change between downstream (high field) and upstream (low field) conditions. These shock crossings are a result of a breathing motion of the shock equilibrium location, initiated by changes in the upstream conditions, which causes the shock to oscillate across the observation point as the spacecraft slowly moves outwards. The separation between the four Cluster spacecraft on this occasion was larger than the typical coherence lengths in this region, which makes it impossible to infer the specific shock velocities during this time period, but they are typically expected to be in the $35$ km/s regime \citep{horbury2002, maksimovic2003}. The Cluster data also provide estimates of the upstream velocity, $V_u \approx 350$ km/s, number density, $n_u \approx 3$ cm$^{-3}$, magnetic field, $B_u \approx 4$ nT, and Mach number, $M_A \approx 7$. However, it is likely that the actual upstream density is higher as the magnetospheric sampling mode on the ion instrument does not always adequately capture the solar wind beam, and measurements with the solar wind mode approximately 1.5 hours after the outermost bow shock crossing rather suggests that $n_u \approx 9$ cm$^{-3}$, and thus $M_A \approx 12$.  The shock normal is estimated to $\bar{n}\approx[.98,-.11,.12]$ in geocentric solar ecliptic coordinates (GSE), using the \citet{slavin1981} bow shock model with the stand-off distance adjusted to match the Cluster location. (In the GSE system, the X-component is toward the Sun, the Z-component is positive towards north, perpendicular to the Earth's orbital plane, and the Y component completes the right-handed XYZ-system.)  The estimated shock normal gives an angle between the upstream flow vector and the shock normal of $\theta_{Vn}\approx10^\circ$. The upstream wave field observed at this occasion was also typical for that observed at similar shocks, with the main ULF power in the 0.1 Hz frequency range, with  wave amplitudes relative to the background field of $\Delta B/B \approx 1$, and occasional bursts of higher-frequency ($\sim$ 1 Hz) whistler waves. A few large-amplitude waves are present in the upstream data, but no signatures typical of SLAMS.  

Reflection events can be identified using data from the Hot Ion Analyser (HIA) of the Cluster Ion Spectrometry (CIS) experiment \citep{reme1997}, which provides 3-D velocity space coverage of the ion distribution function at a 4~s sampling rate. The HIA data does not provide species resolution, and all counts are assumed due to protons. The CIS HIA ion data are collected in a spherical spacecraft-centered coordinate system as a function of ion energy, elevation and azimuth. For the data presented here, the azimuthal component is acquired with a 22.5$^\circ$ resolution over the spacecraft spin period, and the elevation component is similarly resolved in 22.5$^\circ$ steps. The resulting ion velocity distribution provides a much improved velocity space resolution compared to those previously obtained by the ISEE spacecraft \citep[e.g.,][]{onsager1990}, which were limited to collapsed 2-D projections of the ion distribution.

In order to simplify the data analysis, we convert these spacecraft centered velocities into a high-resolution Cartesian grid, using a nearest-neighbor interpolation scheme. This treatment enables arbitrary 2-D plane cuts to be taken through the velocity distribution, while preserving the energy, azimuth and elevation dependence of the sampling points. An example of the C1 CIS ion velocity space data is given in Figure 2, taken at 18:46:40.  The data is displayed in magnetic field-aligned coordinates with its origin in the spacecraft frame, where $v_{\parallel}$ is defined in the direction of the mean magnetic field over the sampling period, $v_{\perp1}$ is orthogonal to $v_{\parallel}$ and the Sun-Earth line, and $v_{\perp2}$ completes the right-handed coordinate triad. Three planar cuts are taken across $v_{\parallel}$, $v_{\perp1}$, and $v_{\perp2}$, all intersecting at the center of mass point, which defines a frame where the convective electric field is zero. The gyro-plane of the reflected ions can be determined from Eqs.~(1) and~(2), assuming an upstream flow velocity of 350 km/s. Although a precise determination of the point in velocity space corresponding to specular reflection requires knowledge of the shock rest frame, we can approximate this with the spacecraft frame, which is sufficiently close for our purposes assuming typical shock velocities.  The location of the spacecraft/shock plane and the reflected ion gyro-plane is indicated by the dotted and dashed lines in Figure~2, respectively.  Figure~3 shows the time evolution of the ion phase space distribution function during a short period close to a shock transition. To aid in the interpretation of the 3-D velocity space data, which is generally difficult to visualize, Figure~3 provides 2-D cuts of four velocity planes of interest, taken at three consecutive sampling periods. The velocity space cut-planes are the following: P1 is the $v_{\parallel}-v_{\perp1}$ plane containing the $v_{\perp2}$ center of mass; P2 is the $v_{\parallel}-v_{\perp2}$ plane containing the $v_{\perp1}$ center of mass; P3 and P4 are $v_{\perp1}-v_{\perp2}$ planes at the shock velocity ($v_{\parallel}=0$) and at the parallel velocity predicted for specularly reflected ions, respectively. The planes P1--4 are labelled in Figure~2 for reference.

In the reflected ion gyro-plane, the expected trajectory of the reflected ions is marked by the black circle, with the black dot indicating the initial velocity space point of reflection. Ion gyration is in the clockwise sense. The three times shown in Figure 3 show typical ion velocity signatures. The first sample, taken at 18:46:40 (row A: P1-P4), shows a clear localized enhancement in the ion phase space density close to the expected reflection point. This gives a strong indication of a reflected ion beam which has yet to spread in velocity space. Such cold ion beams can be observed at other times during the period shown in Figure 1. In all cases, there is also higher energy ions present at lower phase space density, but here we concentrate on the regions of highest value. The second interval (row B) shows a velocity distribution that is patchy and difficult to interpret, and this behavior can typically be associated with a dispersed and time aliased distribution, due for example to changes in the magnetic field direction or the ion velocity distribution during the measurement period. The third sampling interval (row C) shows a velocity space signature that is remarkably different from the cold beam shown in row A, but still consistent with ion reflection. In the reflected velocity plane, there is a clear enhancement present in the ion phase space density, starting at the initial point of reflection, but covering almost half the gyration path. This population is also present over a much broader interval in $v_{\parallel}$, from the shock frame up to velocities exceeding that expected for specular reflection. This is indicative of a much more dispersive reflection process than the cold ion reflection discussed earlier. This type of signature is relatively common in the vicinity upstream of a shock transition, and we will argue that it is central to understanding the ion injection process. This is explored in the next section with the help of a hybrid simulation with parameters similar to the observed shock transition.

\section{Simulation Environment}

The simulation is performed using the hybrid code HYPSI, where the ions are treated as particles and the electrons are considered as a charge-neutralizing massless fluid, with the electron inertial and kinetic effects assumed negligible. The electric and magnetic fields and the ion current density are advanced using the CAM-CL algorithm (current advance method and cyclic leapfrog, see \citet{matthews1994}). The simulation is performed in a two-dimensional rectangular domain in space, 3200 $\lambda_\mathrm{i}$ in X and 128 $\lambda_\mathrm{i}$ in Y, with a cell size of 0.5 by 0.5 $\lambda_\mathrm{i}$. The velocity, magnetic field and electric field vectors are all three-dimensional. Simulation parameters are normalized to the upstream background ion number density, magnetic field, and the proton ion inertial length, $\lambda_\mathrm{i} = V_A/\Omega_\mathrm{i}$, where $\Omega_\mathrm{i}$ is the ion cyclotron frequency. 

Plasma is injected with a super-Alfv\'{e}nic velocity at the left-hand side of the domain though an open boundary which allows flow in both directions. The right-hand boundary acts as an impermeable, reflecting wall, and the domain is periodic in the Y direction. This setup, which is commonly used for simulating shocks, initially creates a backflow of plasma moving in the negative X direction, away from the right-hand boundary. This process rapidly initiates a shock between the inflowing and reflected plasma populations, moving away from the reflecting wall. Ions energized at the shock in the early stages of the simulation produce upstream waves which are convected towards the shock and which in turn affect the shock structure. With time these processes achieve a self-consistent solution which has a steady average shock speed, although the details of the shock transition are highly dynamic. 
%
A finite low-valued resistivity, $\eta=10^{-3}$ $\omega_\mathrm{pi}^{-1}$, is used in order to suppress very short wavelength fluctuations that are otherwise undamped in the hybrid system. The chosen value leads to realistic magnetic fluctuation levels when compared to the observations, and it is a value typical of other studies of shocks and turbulence.
The simulation uses 100 ions per cell for the upstream flow, and a time step of $5\times10^{-3}$ $\Omega_\mathrm{i}^{-1}$. The plasma inflow speed is 6 $V_A$, resulting in a shock return velocity of -2.1 $V_A$ in the simulation frame, and thus a Mach number of $M_A=8.1$ at the shock front. 

The upstream ion population has an isotropic Maxwellian velocity distribution, with a $\beta_\mathrm{i}=0.5$. The magnetic field is initialized at an angle of 30$^\circ$ to X, with $B_x=0.87$, $B_y=0.5$, and $B_z=0$. It should be noted that we do not impose any initial upstream turbulence or wave perturbations during the shock initiation phase, but we follow the shock until a self-generated, fully developed upstream wave field has been developed. As is well known for electromagnetic ion beam instabilities, the most unstable modes propagate parallel to the magnetic field. The inclination of the magnetic field to the shock surface thus produces angled wave fronts in the upstream region, these waves develop slowly over the initial phase of the simulation, and it is not until $t\approx$200 $\Omega_\mathrm{i}^{-1}$ that they are properly established throughout the foreshock. We find that it is important that the scale size in the Y-direction is adequately large in order for these waves to develop properly, as the periodic boundary conditions in Y will constrain the wave modes that can develop as a function of phase velocity, propagation angle, and wavelength. Too narrow a simulation domain in the Y direction typically produces an upstream wave field in which the wave normals are mostly parallel to the shock normal in the $-X$ direction.

Figure~4 shows a snapshot of the simulation taken at $t$=268 $\Omega_\mathrm{i}^{-1}$ with the magnetic field magnitude overlaid with in-plane magnetic field, a profile of the magnetic field magnitude at a fixed value of Y, and the local value of the angle between the magnetic field and the $-X$ direction, $\theta_{Bn}$, again with the magnetic field projection overlaid. The shock is highly dynamic and a video covering the time period analyzed is provided as auxiliary material. Figure~4 shows a clear shock front at $X \approx$ 2675 $\lambda_\mathrm{i}$, with a maximum compression of the magnetic field of a factor of~5. The upstream waves are most easily identified in the $\theta_{Bn}$ view, which shows a variation between almost perfectly parallel and perpendicular geometries in the upstream region. The upstream wave profile is qualitatively similar to that of the observations, which is dominated by long-period ULF waves, often with a steepened trailing edge intertwined with short bursts of higher-frequency whistler waves. The magnetic field compression associated with the ULF waves in the simulation is typically $\sim$1.5, which is in the right range, but slightly lower than the corresponding values in the observations ($\sim$2).

In the simulation, full ion velocity data is recorded at intervals of 0.75 $\Omega_\mathrm{i}^{-1}$ between $t$=250--300 $\Omega_\mathrm{i}^{-1}$.  The high sampling rate for the particle data gives the possibility of a detailed dissection of the ion dynamics at the shock front. This will be used to identify injection densities in Section~4 below.

The simulation is intended to capture the initial ion energization at the shock, but not diffusive shock acceleration in itself; in order to adequate resolve this process, this would require much larger spatial scales and a much longer run time. This scenario is applicable to the terrestrial bow shock, where the energetic upstream ions only reach moderate energies, and have little effect on the shock itself. The bow shock is a small system so that energy gains are limited and the energetic population is strongly influenced by losses and temporal evolution effects. This should be taken into account when making comparisons with other astrophysical shocks.

\section{Ion Injection}

With the high sampling frequency and the large number of particles traced in the simulation, we can resolve both the ion dynamics of the shock as well as the eventual fate of any reflected ions. This data set is also sufficient to resolve the ion velocity distribution on the spatial and temporal scales needed to identify both cold and dispersed ion reflection signatures, and to investigate their origin. One of the main difficulties in the analysis is to know when and where these reflection events are taking place. As there is strong evidence that the specular reflection of ions at the quasi-parallel shock is the initial step toward ion injection, for this reason we will let the ``injected'' particles (i.e., those extracted from the thermal distribution to reach high energy) guide our search for reflection events. The aim of the following section is to provide an overview of the injected particles and to map out where in space and time they first interact with the shock. This data will act as a guide to the event selection in the reflection analysis. 

In order for a particle to be injected this requires an energy of ~5-10 times the shock ram energy $E_{sh}= m_\mathrm{i}(M_A V_A)^2/2$ \citep{caprioli2015}. 
%
Injection is typically supressed for quasi-perpendicular shocks, where the ion acceleration is restricted to shock drift acceleration \citep[e.g.][]{caprioli2014-I}.
We can estimate the injected ion population in the simulation by selecting ions with a velocity exceeding 15 $V_A$ at the final time step of the simulation. This level is lower than that suggested by \citet{caprioli2015}, but this is to account for the finite time required for the energization: the lower threshold leads to better particle statistics in the final time steps. 
%
%
The implications of this threshold will be discussed later in this section.

For each of these ions, we identify their initial point of interaction with the shock. If this takes place as a specular reflection, as postulated above, this classification can simply be achieved by tracking the $v_x$ velocity of the ion, which is expected to be 6 $V_A$ in the upstream region (in the simulation frame) and to become negative at the shock interaction. Particles which do not show typical solar wind energies during the first three time steps with full particle data are excluded from the search, in order to remove upstream ions that have been energized in the earlier stages of the simulation. With the first reversal in $v_x$ signifying the ion's first interaction with the shock front, we can use this information to determine the injection rate of the ions as a function of time and space.  We will hereon refer to these initial velocity reversals as ion injection events for simplicity, however, the injection process should be considered as the whole energization process that takes place over the particles first few interactions with the shock, while it is still contained within approximately a gyroradius of the shock front. 
%
Diffusive shock acceleration will be required in order for the ions to reach an energy span much beyond the injection energy. Limited diffusion can still take place in the upstream region of the simulation, and the most energetic ions reach velocities of 40-50 $V_A$ (equivalent to an ion energy on the order of 20 keV). However, some of the ions that reach sufficient energy to escape upstream will be lost at the upstream boundary of the simulation, thereby artificially limiting the high-energy tail of the energized particle population.

Figure~5 shows contours of the injected ion density at time step $t$=277.75 $\Omega_\mathrm{i}^{-1}$ as a function of X and Y, overlaid on the magnetic field (upper panel) and $\theta_{Bn}$ (lower panel). We can see from the figure that the injection density at this time step is limited to a few regions in the spatial domain, and that these injection points coincide very well with the contours of both $B$ and $\theta_{Bn}$. This shows that the method used for determining the ion injection location is reliable. The $\theta_{Bn}$ panel also shows the impact of a ULF wave front at the shock at $X \sim$ 2650 $\lambda_\mathrm{i}$  and $Y\sim$60--90 $\lambda_\mathrm{i}$. This ULF wave is characterized by a tongue of high $\theta_{Bn}$ extending into the upstream medium, and as it propagates it leaves behind a region of the shock where the magnetic field downstream is strongly perpendicular, the upstream magnetic field is nearly parallel, and there is a sharp transition from low to high field values at the shock front. This configuration enhances particle injection, as it enables an almost ideal specular reflection of incoming ions at the shock front which are then free to escape into the upstream region along the magnetic field. This leads us to believe that the particle injection occurs in spatially localized regions of the shock, and that these regions are specifically characterized by where the sharp magnetic field gradients associated with the quasi-perpendicular shock interact with the trailing edges of the large amplitude ULF waves, where the magnetic field orientation changes abruptly from a perpendicular to a parallel configuration. 

These speculations are confirmed by Figure~6, which shows the injected particle density (black contours) and the local $\theta_{Bn}$ at the shock as a function of time and Y.  The obliquely oriented wave fronts incident on the shock lead to a motion of the pattern of near-upstream $\theta_{Bn}$ in the positive Y direction as a function of time, with the high--low $\theta_{Bn}$ boundary defined by where the trailing edge of this wave intersects the shock front. This motion is sometimes discontinuous as the shock stalls and reconfigures, but the shift toward positive Y with time is generally valid throughout the simulation. The ion injection contours, which are shown in black, are all located in regions where the local shock normal is quasi-parallel, almost immediately following the quasi-perpendicular to quasi-parallel transition. These results are in very good agreement with the conclusions from the initial analysis of the time step shown in Figure~5.  In addition to the spatial overview, Figure~6 also shows an estimate of the percentage of injected particles as a function of time, integrated along the entire length of the shock. This trace shows that injection typically occurs in bursts, strongly dependent on the local configuration of the shock. The injection rate averages to $\sim 0.35\%$ over time and space, but local injection rates exceeding $3\%$ are frequently observed. 
%
%
These levels are only indicative of the true injection rate, as they depend on the definition of the injection energy threshold. The true ratio of ions that indeed go into diffusive shock acceleration is uncertain, partly due to the finite time required for the acceleration process. On average, the acceleration from solar wind velocities up to 15 $V_A$ is achieved in $\sim$15-30 $\Omega_\mathrm{i}^{-1}$, 18 $V_A$ in $\sim$20-40 $\Omega_\mathrm{i}^{-1}$ and 25 $V_A$ in 40+  $\Omega_\mathrm{i}^{-1}$. Increasing the energy threshold to 18 $V_A$ (5 $E_{sh}$) or 25 $V_A$ (10 $E_{sh}$) leads to a ~10\% and ~50\% reduction in the number of injected ions in the first half of the time interval analyzed, respectively, but the spatial pattern in Figure~6 remains the same. This indicates that the injection rates given here overestimate the number of ions that go into diffusive shock acceleration by a factor of two at most.  
These numbers are also in line with previous simulations by \citet{caprioli2014-I}, who report an injection rate of $10^{-4}$--$10^{-3}$ for quasi-parallel shocks. Knowing that the initial ion injection is variable in time and constrained to narrow regions in space, we now focus the search for energized ion velocity distributions to the region upstream of such injection events.

Figure~7 shows an ion distribution from the simulation in panel~A, taken from a box close to the shock interface at $t$=278.5 $\Omega_\mathrm{i}^{-1}$, i.e., one time step after the ion injection event shown in Figure 5. The velocity distribution is shown in the format as that given for the Cluster data in Figure 3, with 2-D velocity space cut-planes at the center of mass (P1 and P2), the shock frame (P3), and at the ion reflection velocity gyro-plane (P4). Reflected gyrating ions follow a left-hand trajectory in the reflected plane, however, note that there is a 180$^\circ$ shift in the initial reflection point compared to that given in Figure 3, due a polarity change in the magnetic field direction compared to the shock normal. There is a large population of backstreaming ions in this region that are headed away from the shock. These are visible in both panels P1 and P2, and they show a spread in velocity that ranges from the shock frame velocity to negative parallel velocities exceeding that expected for reflection. This ion distribution also extends to positive velocities, i.e., ions that are currently travelling toward the shock. These backstreaming ions cover approximately half of the gyro-circle in both the shock and reflection planes, consistent with a left-hand (clockwise) gyration from the initial reflection point in velocity space. The velocity signature closely resembles that of the dispersed reflection events in the observations rather than a cold reflection ion beam, and it gives strong evidence that connects the Cluster observations with the ion injection process; the comparative velocity distribution from the observations is shown in panel B, with the coordinate system rotated so that the initial reflection point aligns between the two panels. Although the two velocity space signatures are qualitatively similar, it should be noted that we are only able to extract count rates rather than phase space densities from the simulation at this point. A quantitative comparison would require a much larger number of particles per cell. The main difference between the simulated and observed distributions is the extended contour of the solar wind beam in Panel A:P2, which covers a range of  $v_{\perp2}$ from -5 to +5. This spread in velocity is due to solar wind particles that are beginning to interact with the shock front, being diverted from their original velocity space location by the magnetic field in the shock transition region.

The simulation also provides an opportunity to investigate both the origin and the fate of the ions by tracing their spatial location and velocity space distribution over time. We will show here that the ion acceleration can be understood in terms of the \citet{sugiyama1999} scatter-free ion acceleration model, which is based on phase trapping in large-amplitude monochromatic upstream and downstream waves. A particle gains energy as it moves between the downstream (slowed and compressed) wave and the incoming upstream wave. Within each wave frame, the particle velocity is restricted to a circle in $v_\parallel$--$v_\perp$ space, centered at the wave phase velocity. We will here refer to these circles as iso-energy contours. With the change of wave frame from upstream to downstream, the appropriate iso-energy contour changes, leading to an acceleration of the ion with each frame transition. 

However, the \citet{sugiyama1999} model is restricted in that it only considers the ion motion at parallel shocks. For an oblique shock configuration, and if specular reflection is dominant, these iso-energy contours are not defined by the field-aligned velocity, but rather the shock-normal velocity of the ion, as this determines the ion motion relative to the shock front. In this coordinate system, here represented by $v_x$ and $\sqrt{v_y^2+v_z^2}$, the ion velocities follow the velocity space trajectory expected. This is not always apparent in the individual particle trajectories, as this requires a higher particle sampling rate, but it is evident when the overall distribution of the ions is considered. This is shown in Figures~8 and~9. Both of these figures show the spatial location of the ions in the simulation domain on the left, displayed on top of the magnetic field magnitude, and the velocity distribution of the selected ions on the right. Figure~8 shows three time steps that explain the initial ion interaction with the shock, and Figure~9 shows three time steps later in the simulation that give further information on the injection process and the spatial and velocity space diffusion of the ions over time. The time and space requirements on the ion selection are similar to those in Figure~7, but with an additional selection criterion to separate the ions undergoing injection from the downstream thermal population and those ions that have been energized already in an earlier phase of the simulation.  This constraint is imposed by requiring that the ion velocity is near the upstream velocity in the first recorded time steps, and that the ion is located upstream of the shock at the end of the simulation. The selected ions are marked by yellow circles in the spatial plots. The remaining ions from the selection domain are marked by black dots in the background for comparison. The velocity space distributions shown are for the injected (yellow) population only. These figures also show iso-energy contours in the upstream (black) and the shock frame (green). 

In the first time step of Figure~8 (panel A, $t$=277.75 $\Omega_\mathrm{i}^{-1}$), a large part of the ions are still integrated in the upstream flow, with velocities that are near the upstream velocity, marked by the black dot at the center of the upstream iso-energy contours. These ions are therefore yet to encounter the shock. (Note that the choice of color map means that the peak of the incident distribution is saturated, and thus much larger than it might appear.) Some ions have already interacted with the shock, and their velocities are spread over a higher energy shell in the upstream frame. Most of these ions are headed back upstream, however, some are already turning back toward the shock (identified by $v_x>0$), which means that they will re-encounter the shock with a higher energy than the cold upstream flow. These particles have been reflected at an earlier time, but much less than a cyclotron time before the main bulk of the ion population.  This time aliasing effect on the ion distributions can explain the diffuse nature of the reflected ion beams seen in the Cluster ion velocity measurements. Although the reflection process is near-specular and non-dispersive, small variations in the time and space of the reflection points will naturally create a more diffuse ion population, where some have just been reflected off the shock, and others have been reflected earlier and then turned around by the upstream magnetic field and are headed back toward the shock.

Panel~B shows the ion distribution at $t$=278.5 $\Omega_\mathrm{i}^{-1}$, which is the selection time step. The time and space restrictions on the ion selection are here clearly visible, as all ions can be seen to converge in space to the selection box.  A majority of the ions have now been reflected by the shock at least once, and most ions are found within the 10--15 $V_A$ upstream iso-energy contours. Another important feature is that the part of the velocity distribution of the energized ions that have $v_x>5$ have now clearly diverted from the upstream iso-energy contour to the intersecting shock frame contour. This relates to a second interaction with the shock, which changes the appropriate reference frame of these ions, and it is a step toward further energization.  At the final time step shown in Figure 8 (panel C, $t$=280.00 $\Omega_\mathrm{i}^{-1}$), it is clear how the $v_x>0$ part of the ion distribution follows the shock frame contour (especially visible at the lower threshold), whereas the ions with $v_x<0$, which have completed their second interaction with the shock, are settling at a higher upstream iso-energy compared to the two earlier times. This alternating change of frame between the upstream and downstream domains is what drives the ion injection.  The time steps shown in Figure~9 also highlight these energy frame transitions as the injection process continues, in panel A ($t$=281.5 $\Omega_\mathrm{i}^{-1}$), all of the ions have now reached an energy of $\sim$15 $V_A$ in the upstream frame,  and almost all are headed away from the shock along the magnetic field. As they encounter the next upstream wave front the magnetic field orientation changes from quasi-parallel to quasi-perpendicular (panel B, $t$=283.75 $\Omega_\mathrm{i}^{-1}$), which means that the ions are now swept back toward the shock again with the upstream flow, leading to another interaction with the shock, and the associated transition from upstream to shock energy contours. The ion distribution is now becoming more and more diffuse, both in space and in velocity space. This is particularly visible in the final time step shown (panel C, $t$=295.75 $\Omega_\mathrm{i}^{-1}$), where some ions are caught in the next upstream wave, whereas others are beginning to escape the ULF wave field into the next parallel section in the upstream field, thereby escaping the shock front and completing the injection process.

Finally, some of the requirements for ion injection can be understood by considering the field-aligned velocity of the ions, which is shown in Figure~10. The first panel shows the initial velocity distribution in the upstream region, where the bulk of the ions have positive field-aligned velocities. This changes during the first shock encounter (panel B, $t$=278.5 $\Omega_\mathrm{i}^{-1}$), and all of the injected ions quickly establish a negative $v_\parallel$ (Panel C, $t$ = 280 $\Omega_\mathrm{i}^{-1}$). This process is completed within approximately one ion gyroperiod following reflection, and it is retained until the upstream diffusion begins to play a role. In contrast, the majority of the downstream ions retain a positive $v_\parallel$. This can be understood in connection with the spatial distribution shown in Figures 8 and 9, as the sign of $v_\parallel$ separates the ions that are attempting to head upstream along the magnetic field, despite periodic encounters with the upstream ULF field, from those that are headed downstream. All of these ions are still confined to the same flux tube, and the spread in $v_\parallel$ accounts for the spatial spread along the magnetic field. The right-hand panels of Figure 10, which show the perpendicular velocity distribution of the injected ions, give important clues about the gyrotropy of the ions. It can be seen here that the ions retain some gyrotropy after the first shock reflection (Panel B), however, this is quickly lost as the ions spread out in space and velocity within one ion gyroperiod after reflection (Panel C). Although it should be remembered that the latter distributions do not correspond to what would be observed by a spacecraft since they are accumulated over particles which are spread out in space.    

\section{Discussion and Conclusions}

The data presented here shows how the spacecraft observations and the hybrid simulations can be merged into a complete and validated description of the ion injection process at oblique quasi-parallel shocks. The Cluster measurements, provided in Figure 3, show that the velocity space signatures associated with ion reflection at the quasi-parallel shock often is more complex than that typically associated with cold specular reflection events. In these events, the reflected ions cover a much larger region in velocity space than the cold reflected ion beams, with a spread in parallel velocities that ranges from the shock frame to beyond the reflection plane, and a spread in gyrophase over nearly half the gyro-circle. The same velocity signatures are reproduced by the hybrid simulations, as shown in Figure 7. Here, the simulated ion velocity distribution contains the same spread in gyrophase, starting at the initial point of specular reflection, and a very similar spread in parallel velocities all the way from the shock frame to the reflected ion velocity frame. These dispersed reflection events are characteristic of the ion injection process at quasi-parallel shocks. The velocity space signatures can be understood in terms of specular ion reflection taking place over a small range of locations in both space and time  in a region where the magnetic field is quickly changing configuration, which gives rise to the observed spread in velocity space. The specific procedure for ion injection is summarized in the five points below:
 
\noindent (1) The injected ions first encounter the shock at the trailing edge of a ULF wave, where the shock configuration rapidly changes from quasi-perpendicular to quasi-parallel. In this region, a locally quasi-perpendicular field downstream of the shock allows efficient ion reflection into an upstream field which is quasi-parallel. The reflected ions can subsequently escape into the upstream region, having gained their first important boost in energy.

\noindent (2) At this wave boundary, some ions may undergo a series of rapid reflections, leading to what in the observations may look like a dispersed ion population as they encounter other reflected ions. This is in reality a coherent but time and space dependent process; the spread in ion velocity can be accounted for by small variations in the reflection time and local variations in shock configuration over an ensemble of particles, rather than a purely stochastic process.

\noindent (3) Following reflection, the ions acquire a field-aligned velocity that is directed upstream in the shock frame. This is essential for the subsequent escape from the shock interface, and it leads to the eventual separation of the upstream and downstream populations along the magnetic field.

\noindent (4) As the ions encounter the next wave front in the upstream region, the local field direction changes from quasi-parallel to quasi-perpendicular. The ions are convected back toward the shock with the upstream flow, leading to a reflection at a locally quasi-perpendicular shock. This process can be repeated several times, and at each of these instances, the change in the reference frame from the upstream to the shock frame and back leads to an energization of the ions, even though they are constrained by the iso-energy contours in each respective frame.

\noindent (5) Once the ion gains enough energy in the field-aligned direction to overcome the upstream wave field without any additional interaction with the shock, it is free to escape from the  immediate upstream region of the shock, and the particle injection process is completed.

This description builds on many previous studies of the quasi-parallel shock, combining these into one encompassing model. The importance of specular reflection for ion injection has long been suspected \citep{gosling1982, gosling1989, onsager1990, scholer1990-II}. These results are also consistent with the idea that the quasi-parallel shock is the primary region for ion injection \citep[e.g.,][]{caprioli2015}, and that injection can be triggered by local changes in the magnetic field geometry of the upstream wave field \citep{scholer1990-I, kucharek1991}. That the quasi-parallel shock locally becomes quasi-perpendicular due to the upstream wave field was first shown by \citet{scholer1992}, and that sudden changes in the shock configuration from perpendicular to parallel due to solar wind discontinuities can lead to temporarily enhanced injection rates has been shown by \citet{kucharek1995}. The ion energization is achieved through repeated transitions between the upstream and downstream wave frames, in a matter similar to that presented by \citet{sugiyama1999}. The ions require multiple reflections at the shock before they can be injected, as they need to gain enough energy in the magnetic field-aligned direction to escape the upstream wave field \citep[e.g.,][]{caprioli2015}.  We also find that the ion injection process can be fast; it can be completed within as little three to five ion gyroperiods after the initial shock encounter. These time scales are similar to those expected at near parallel shocks \citet{sugiyama1999}.
%
%
This description of the ion injection process is compatible with the model provided by \citet{caprioli2015}. The ion reflection that initiates the injection process requires the steep shock discontinuity provided by a quasi-perpendicular-like shock jump. These regions occur periodically by the impacting wave fronts, similar to the shock reformation described by \citet{caprioli2015}. Not all ions are reflected in these injection regions, many are still advected downstream contributing to the thermal downstream population. A fraction of the ions are transmitted downstream at subsequent shock encounters,  having gained energies intermediate between the thermal and injected populations, and shock-drift accelerated ions can also be generated at locally quasi-perpendicular sections of the shock. These will be contribute to a supra-thermal ion population, as identified by \citet{caprioli2014-I}. These findings are also in agreement with previous observation of two alternating states in the thermal distribution of the magnetosheath ions downstream of the terrestrial bow shock reported by \citet{thomsen1990}, where cool and hot ion populations were periodically observed. 
  
The main purpose of this paper has been to present a consolidated analysis of hybrid simulations and spacecraft observations. The good agreement between the modelled and observed velocity distributions shows that the hybrid model can provide a reliable description of the ion dynamics at collisionless shocks, and the important role played by the upstream wave field. It is important to note that the success of the comparison validates the hybrid method for studies of ion acceleration at most heliospheric shocks. For the reason of the spacecraft comparison, this study is naturally constrained to the set of shock parameters specific for this event; the model should be applicable for a range shock obliquities and Mach numbers, but a proper parameter survey will be required to determine quantitative estimates on the injection rate the different parameter regimes. Further work also includes analysis of the velocity space signature in a range of different regions, both upstream and downstream of the shock.

\acknowledgments
We thank Cluster FGM and CIS instrument teams and the Cluster Science Archive for providing the Cluster data used in this study, which are publicly available at http://www.cosmos.esa.int/web/csa. We are grateful to Manfred Scholer for useful comments on the manuscript. DB and CXM acknowledge support from ISSI (Bern, Switzerland) for the science team ``Physics of the Injection of Particle Acceleration at Astrophysical, Heliospheric, and Laboratory Collisionless Shocks.'' The research leading to the presented results has received funding from the European Commission's Seventh Framework Programme FP7 under the grant agreement SHOCK (project number 284515), and the work was also supported by the UK Science and Technology Facilities Council (STFC) grant ST/J001546/1. This work used the DiRAC Complexity system, operated by the University of Leicester IT Services, which forms part of the STFC DiRAC HPC Facility (www.dirac.ac.uk ). This equipment is funded by BIS National E-Infrastructure capital grant ST/K000373/1 and STFC DiRAC Operations grant ST/K0003259/1. DiRAC is part of the National E-Infrastructure.


\bibliographystyle{apj}
\bibliography{shockref}{}

\begin{thebibliography}{}
\expandafter\ifx\csname natexlab\endcsname\relax\def\natexlab#1{#1}\fi

\bibitem[{Armstrong {et~al.}(1985)Armstrong, Pesses, \& Decker}]{armstrong1985}
Armstrong, T.~P., Pesses, M.~E., \& Decker, R.~B. 1985, in Collisionless Shocks
  in the Heliosphere: Reviews of Current Research, Vol.~35 (AGU Washington,
  DC), 271--285

\bibitem[{{Axford} {et~al.}(1977){Axford}, {Leer}, \& {Skadron}}]{axford1977}
{Axford}, W.~I., {Leer}, E., \& {Skadron}, G. 1977, International Cosmic Ray
  Conference, 11, 132

\bibitem[{{Bell}(1978{\natexlab{a}})}]{bell1978-I}
{Bell}, A.~R. 1978{\natexlab{a}}, \mnras, 182, 147

\bibitem[{{Bell}(1978{\natexlab{b}})}]{bell1978-II}
---. 1978{\natexlab{b}}, \mnras, 182, 443

\bibitem[{Berezhko(2005)}]{berezhko2005}
Berezhko, E. 2005, Advances in Space Research, 35, 1031

\bibitem[{Berezhko \& V{\"o}lk(2007)}]{berezhko2007}
Berezhko, E., \& V{\"o}lk, H. 2007, The Astrophysical Journal Letters, 661,
  L175

\bibitem[{Blandford \& Ostriker(1978)}]{blandford1978}
Blandford, R.~D., \& Ostriker, J.~P. 1978, The Astrophysical Journal, 221, L29

\bibitem[{Blasi(2013)}]{blasi2013}
Blasi, P. 2013, The Astronomy and Astrophysics Review, 21, 1

\bibitem[{Burgess(1987)}]{burgess1987}
Burgess, D. 1987, Journal of Geophysical Research: Space Physics, 92, 1119

\bibitem[{Burgess(1997)}]{burgess1997}
---. 1997, Advances in Space Research, 20, 673

\bibitem[{Caprioli {et~al.}(2015)Caprioli, Pop, \& Spitkovsky}]{caprioli2015}
Caprioli, D., Pop, A.-R., \& Spitkovsky, A. 2015, The Astrophysical Journal
  Letters, 798, L28

\bibitem[{Caprioli \& Spitkovsky(2014)}]{caprioli2014-I}
Caprioli, D., \& Spitkovsky, A. 2014, The Astrophysical Journal, 783, 91

\bibitem[{Decker(1988)}]{decker1988}
Decker, R.~B. 1988, Space Science Reviews, 48, 195

\bibitem[{{Drury}(1983)}]{drury1983}
{Drury}, L.~O. 1983, Reports on Progress in Physics, 46, 973

\bibitem[{{Eastwood} {et~al.}(2005){Eastwood}, {Lucek}, {Mazelle}, {Meziane},
  {Narita}, {Pickett}, \& {Treumann}}]{eastwood2005}
{Eastwood}, J.~P., {Lucek}, E.~A., {Mazelle}, C., {et~al.} 2005, \ssr, 118, 41

\bibitem[{Fairfield(1974)}]{fairfield1974}
Fairfield, D.~H. 1974, Journal of Geophysical Research, 79, 1368

\bibitem[{Gargat{\'e} \& Spitkovsky(2012)}]{gargate2012}
Gargat{\'e}, L., \& Spitkovsky, A. 2012, The Astrophysical Journal, 744, 67

\bibitem[{Giacalone(2004)}]{giacalone2004}
Giacalone, J. 2004, The Astrophysical Journal, 609, 452

\bibitem[{Giacalone {et~al.}(1992)Giacalone, Burgess, Schwartz, \&
  Ellison}]{giacalone1992}
Giacalone, J., Burgess, D., Schwartz, S., \& Ellison, D.~C. 1992, Geophysical
  research letters, 19, 433

\bibitem[{Giacalone {et~al.}(1997)Giacalone, Burgess, Schwartz, Ellison, \&
  Bennett}]{giacalone1997}
Giacalone, J., Burgess, D., Schwartz, S., Ellison, D.~C., \& Bennett, L. 1997,
  Journal of Geophysical Research: Space Physics (1978--2012), 102, 19789

\bibitem[{Gosling \& Robson(1985)}]{gosling1985}
Gosling, J., \& Robson, A. 1985, Collisionless shocks in the heliosphere:
  reviews of current research, Geophys. Monogr. Ser, 35, 141

\bibitem[{Gosling {et~al.}(1982)Gosling, Thomsen, Bame, Feldman, Paschmann, \&
  Sckopke}]{gosling1982}
Gosling, J., Thomsen, M., Bame, S., {et~al.} 1982, Geophysical Research
  Letters, 9, 1333

\bibitem[{{Gosling} {et~al.}(1989){Gosling}, {Thomsen}, {Bame}, \&
  {Russell}}]{gosling1989}
{Gosling}, J.~T., {Thomsen}, M.~F., {Bame}, S.~J., \& {Russell}, C.~T. 1989,
  \jgr, 94, 10027

\bibitem[{Greenstadt {et~al.}(1995)Greenstadt, Le, \&
  Strangeway}]{greenstadt1995}
Greenstadt, E., Le, G., \& Strangeway, R. 1995, Advances in Space Research, 15,
  71

\bibitem[{Guo \& Giacalone(2013)}]{guo2013}
Guo, F., \& Giacalone, J. 2013, The Astrophysical Journal, 773, 158

\bibitem[{Horbury {et~al.}(2002)Horbury, Cargill, Lucek, Eastwood, Balogh,
  Dunlop, Fornacon, \& Georgescu}]{horbury2002}
Horbury, T., Cargill, P., Lucek, E., {et~al.} 2002, Journal of Geophysical
  Research: Space Physics (1978--2012), 107, SSH

\bibitem[{Jones \& Ellison(1991)}]{jones1991}
Jones, F.~C., \& Ellison, D.~C. 1991, Space Science Reviews, 58, 259

\bibitem[{{Krymskii}(1977)}]{krymskii1977}
{Krymskii}, G.~F. 1977, Akademiia Nauk SSSR Doklady, 234, 1306

\bibitem[{Kucharek \& Scholer(1991)}]{kucharek1991}
Kucharek, H., \& Scholer, M. 1991, Journal of Geophysical Research: Space
  Physics (1978--2012), 96, 21195

\bibitem[{Kucharek \& Scholer(1995)}]{kucharek1995}
---. 1995, Advances in Space Research, 15, 171

\bibitem[{Kuramitsu \& Krasnoselskikh(2005{\natexlab{a}})}]{kuramitsu2005}
Kuramitsu, Y., \& Krasnoselskikh, V. 2005{\natexlab{a}}, Physical review
  letters, 94, 031102

\bibitem[{Kuramitsu \& Krasnoselskikh(2005{\natexlab{b}})}]{kuramitsu2005-I}
---. 2005{\natexlab{b}}, Physical review letters, 94, 031102

\bibitem[{Kuramitsu \& Krasnoselskikh(2005{\natexlab{c}})}]{kuramitsu2005-II}
---. 2005{\natexlab{c}}, Journal of Geophysical Research: Space Physics
  (1978--2012), 110

\bibitem[{Le {et~al.}(1992)Le, Russell, Thomsen, \& Gosling}]{le1992}
Le, G., Russell, C., Thomsen, M., \& Gosling, J. 1992, Journal of Geophysical
  Research: Space Physics (1978--2012), 97, 2917

\bibitem[{Leroy {et~al.}(1981)Leroy, Goodrich, Winske, Wu, \&
  Papadopoulos}]{leroy1981}
Leroy, M., Goodrich, C., Winske, D., Wu, C., \& Papadopoulos, K. 1981,
  Geophysical Research Letters, 8, 1269

\bibitem[{Leroy {et~al.}(1982)Leroy, Winske, Goodrich, Wu, \&
  Papadopoulos}]{leroy1982}
Leroy, M., Winske, D., Goodrich, C., Wu, C., \& Papadopoulos, K. 1982, Journal
  of Geophysical Research: Space Physics (1978--2012), 87, 5081

\bibitem[{Maksimovic {et~al.}(2003)Maksimovic, Bale, Horbury, \&
  Andr{\'e}}]{maksimovic2003}
Maksimovic, M., Bale, S., Horbury, T., \& Andr{\'e}, M. 2003, Geophysical
  research letters, 30

\bibitem[{Matthews(1994)}]{matthews1994}
Matthews, A.~P. 1994, Journal of Computational Physics, 112, 102

\bibitem[{Mazelle {et~al.}(2003)Mazelle, Meziane, LeQu{\'e}au, Wilber,
  Eastwood, Reme, Sauvaud, Bosqued, Dandouras, McCarthy,
  {et~al.}}]{mazelle2003}
Mazelle, C., Meziane, K., LeQu{\'e}au, D., {et~al.} 2003, Planetary and Space
  Science, 51, 785

\bibitem[{Onsager {et~al.}(1990)Onsager, Thomsen, Gosling, Bame, \&
  Russell}]{onsager1990}
Onsager, T., Thomsen, M., Gosling, J., Bame, S., \& Russell, C. 1990, Journal
  of Geophysical Research: Space Physics (1978--2012), 95, 2261

\bibitem[{Ptuskin {et~al.}(2010)Ptuskin, Zirakashvili, \& Seo}]{ptuskin2010}
Ptuskin, V., Zirakashvili, V., \& Seo, E.-S. 2010, The Astrophysical Journal,
  718, 31

\bibitem[{{Reme} {et~al.}(1997){Reme}, {Bosqued}, {Sauvaud}, {Cros},
  {Dandouras}, {Aoustin}, {Bouyssou}, {Camus}, {Cuvilo}, {Martz}, {Medale},
  {Perrier}, {Romefort}, {Rouzaud}, {D`Uston}, {Mobius}, {Crocker}, {Granoff},
  {Kistler}, {Popecki}, {Hovestadt}, {Klecker}, {Paschmann}, {Scholer},
  {Carlson}, {Curtis}, {Lin}, {McFadden}, {Formisano}, {Amata},
  {Bavassano-Cattaneo}, {Baldetti}, {Belluci}, {Bruno}, {Chionchio}, {di
  Lellis}, {Shelley}, {Ghielmetti}, {Lennartsson}, {Korth}, {Rosenbauer},
  {Lundin}, {Olsen}, {Parks}, {McCarthy}, \& {Balsiger}}]{reme1997}
{Reme}, H., {Bosqued}, J.~M., {Sauvaud}, J.~A., {et~al.} 1997, \ssr, 79, 303

\bibitem[{Scholer(1990)}]{scholer1990-I}
Scholer, M. 1990, Geophysical Research Letters, 17, 1821

\bibitem[{Scholer \& Burgess(1992)}]{scholer1992}
Scholer, M., \& Burgess, D. 1992, Journal of Geophysical Research: Space
  Physics (1978--2012), 97, 8319

\bibitem[{Scholer \& Terasawa(1990)}]{scholer1990-II}
Scholer, M., \& Terasawa, T. 1990, Geophysical research letters, 17, 119

\bibitem[{Schure {et~al.}(2012)Schure, Bell, Drury, \& Bykov}]{schure2012}
Schure, K., Bell, A., Drury, L., \& Bykov, A. 2012, Space science reviews, 173,
  491

\bibitem[{Schwartz(1991)}]{schwartz1991-I}
Schwartz, S.~J. 1991, Advances in Space Research, 11, 231

\bibitem[{Schwartz \& Burgess(1991)}]{schwartz1991-II}
Schwartz, S.~J., \& Burgess, D. 1991, Geophysical Research Letters, 18, 373

\bibitem[{Schwartz {et~al.}(1992)Schwartz, Burgess, Wilkinson, Kessel, Dunlop,
  \& L{\"u}hr}]{schwartz1992}
Schwartz, S.~J., Burgess, D., Wilkinson, W.~P., {et~al.} 1992, Journal of
  Geophysical Research: Space Physics (1978--2012), 97, 4209

\bibitem[{Slavin \& Holzer(1981)}]{slavin1981}
Slavin, J.~A., \& Holzer, R.~E. 1981, Journal of Geophysical Research: Space
  Physics (1978--2012), 86, 11401

\bibitem[{Sugiyama \& Terasawa(1999)}]{sugiyama1999}
Sugiyama, T., \& Terasawa, T. 1999, Advances in Space Research, 24, 73

\bibitem[{Thomsen {et~al.}(1990)Thomsen, Gosling, Bame, Onsager, \&
  Russell}]{thomsen1990}
Thomsen, M., Gosling, J., Bame, S., Onsager, T., \& Russell, C. 1990, Journal
  of Geophysical Research, 95, 6363

\end{thebibliography}



\begin{figure}
\includegraphics[angle=0,scale=.50]{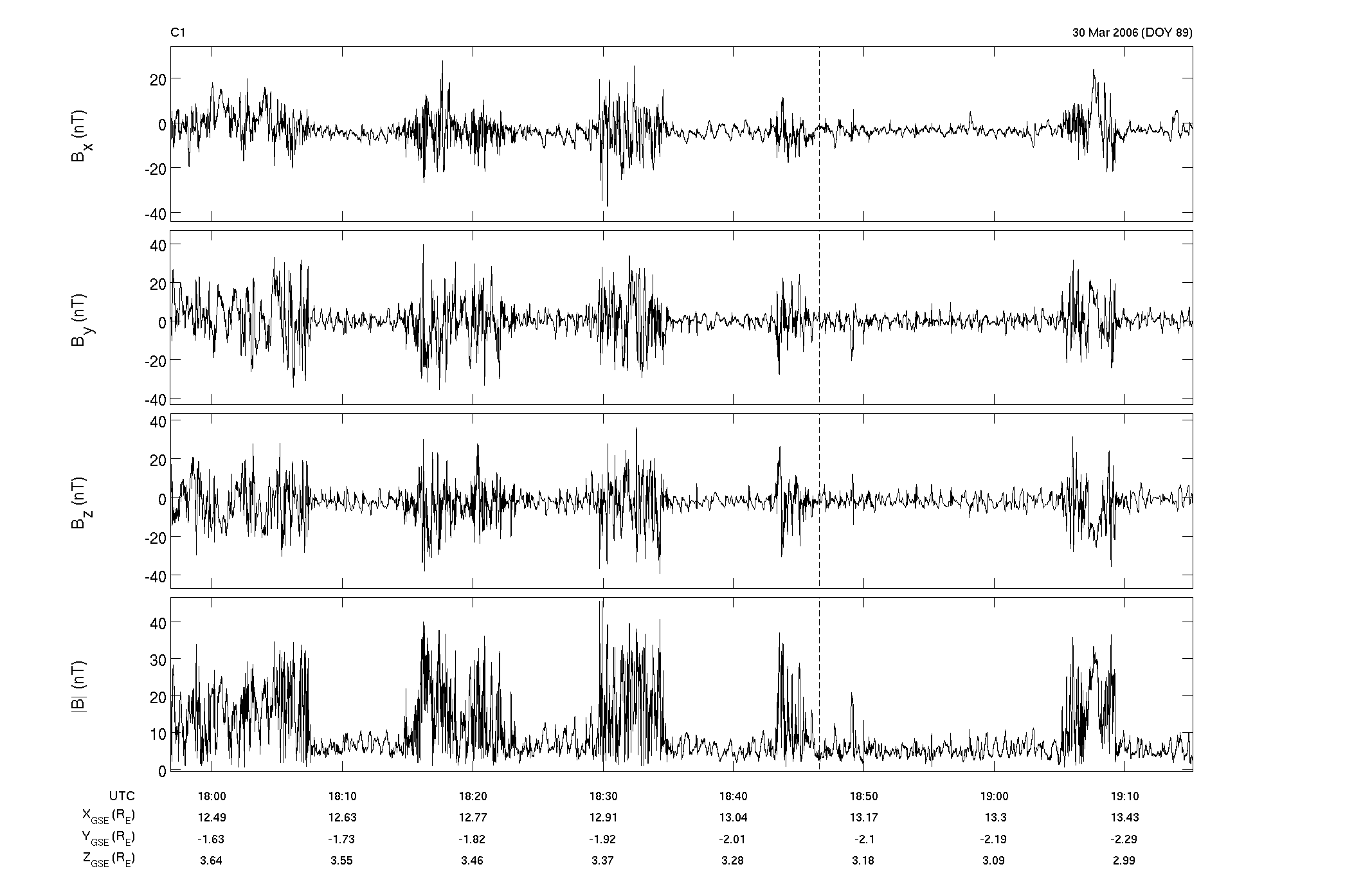}
\caption{Typical magnetic field profile of a quasi-parallel shock crossing, recorded by the Cluster 1 spacecraft. The top three panels show the $X$, $Y$ and $Z$ components of the magnetic field in GSE coordinates, and the bottom panel shows the magnetic field magnitude. The dashed lines mark the time period of the ion spectrum shown in Figures 2 and 3.}
\end{figure}

\begin{figure}
\includegraphics[angle=0,scale=.8]{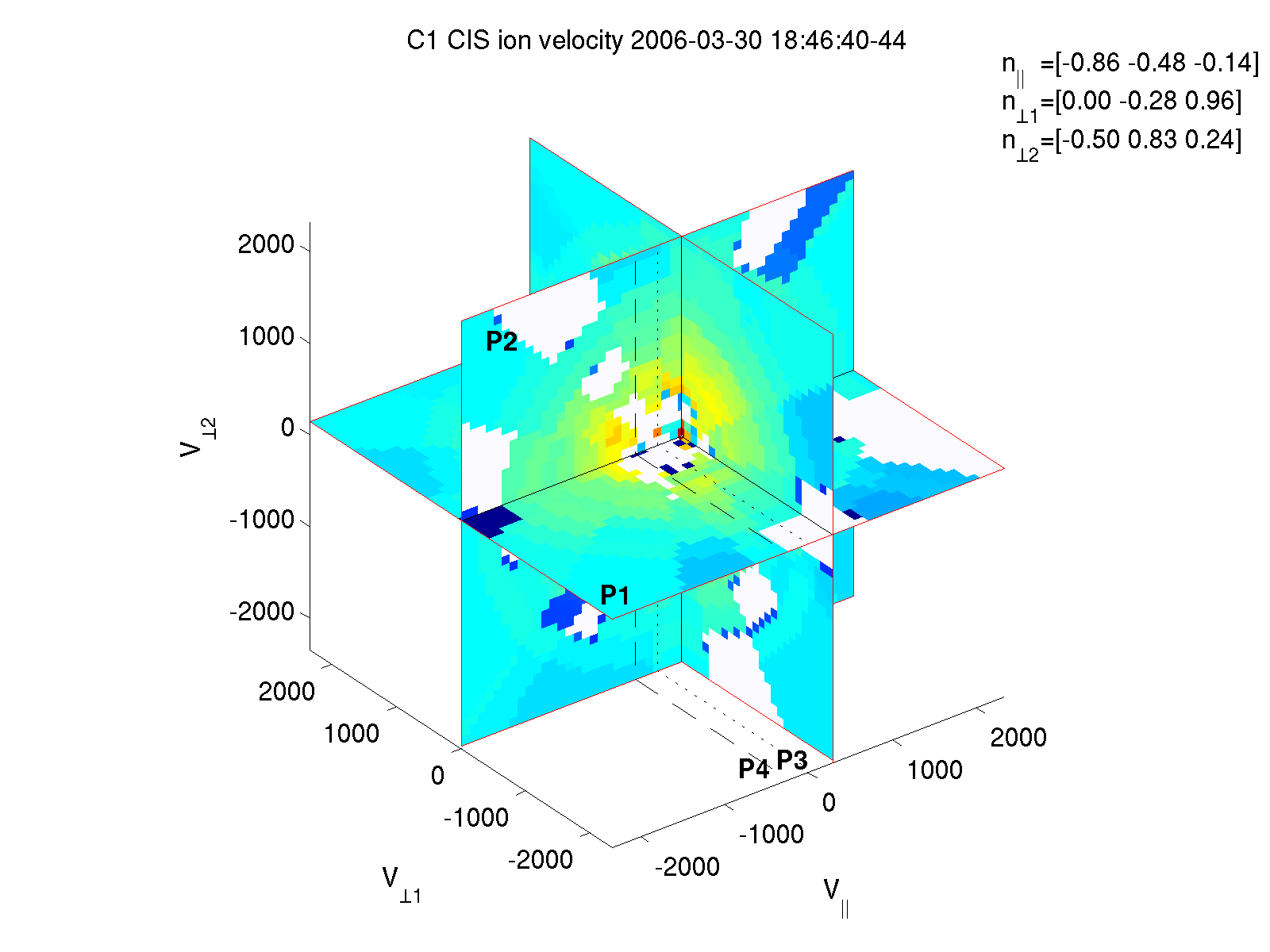}
\caption{Three-dimensional velocity space view of the ion phase space distribution function taken at 18:46:40 UT. The cut-planes are here taken through the center of mass point. The dotted line indicates the location of the spacecraft/shock plane (equivalent to P3 in Figure~3), and the dashed lines the plane in $v_{\parallel}$ expected for perfect specularly reflected ions (P4 in Figure~3).}
\end{figure}

\begin{figure}
\includegraphics[angle=0,scale=.4]{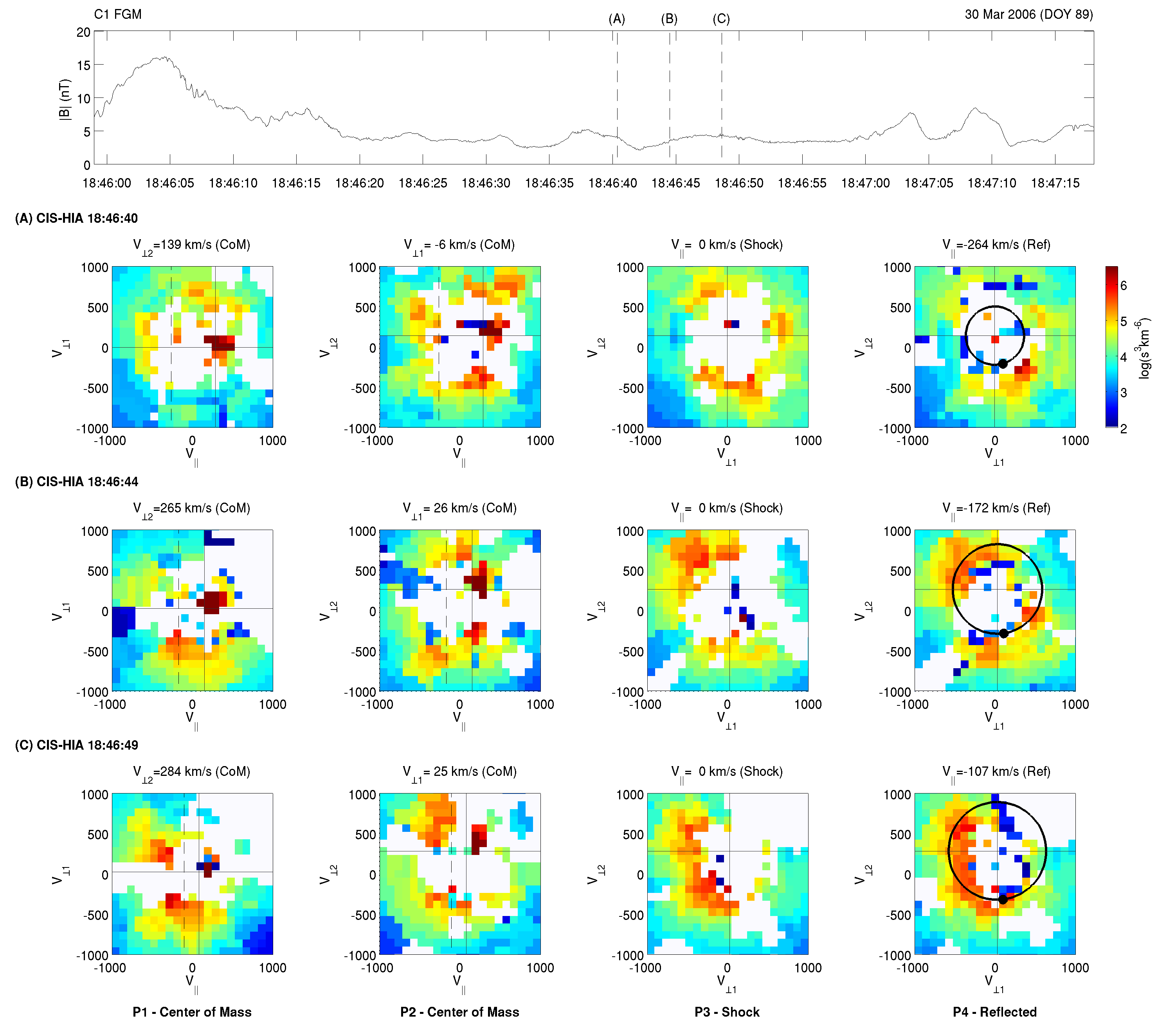}
\caption{Cluster CIS/HIA ion phase space distributions shown as 2-D cut-planes. The top panel shows the magnetic field trace, with dashed lines indicating the center time of the ion spectra. The color panels show cuts of the ion distribution from three consecutive time steps (rows A, B and C), taken at the center of mass (P1 and P2, first two columns of each row), the spacecraft/shock frame (P3), and at the reflection velocity (P4). The solid lines indicate the intersection of the center of mass planes, and the dashed lines indicate the location of the reflection plane.  In the reflection plane, the dot and the circle represents the expected initial reflection point and the gyro-trajectory of the ions, with the ions gyrating in a clockwise sense. An example of cold specularly reflected ion beam is seen in the reflected plane cut in the first time step (i.e., row A, panel P4), and the signature of a more dispersed ion reflection event is seen in the third time step (row A, panels P3 and P4).}
\end{figure}

\begin{figure}
\includegraphics[angle=0,scale=.5]{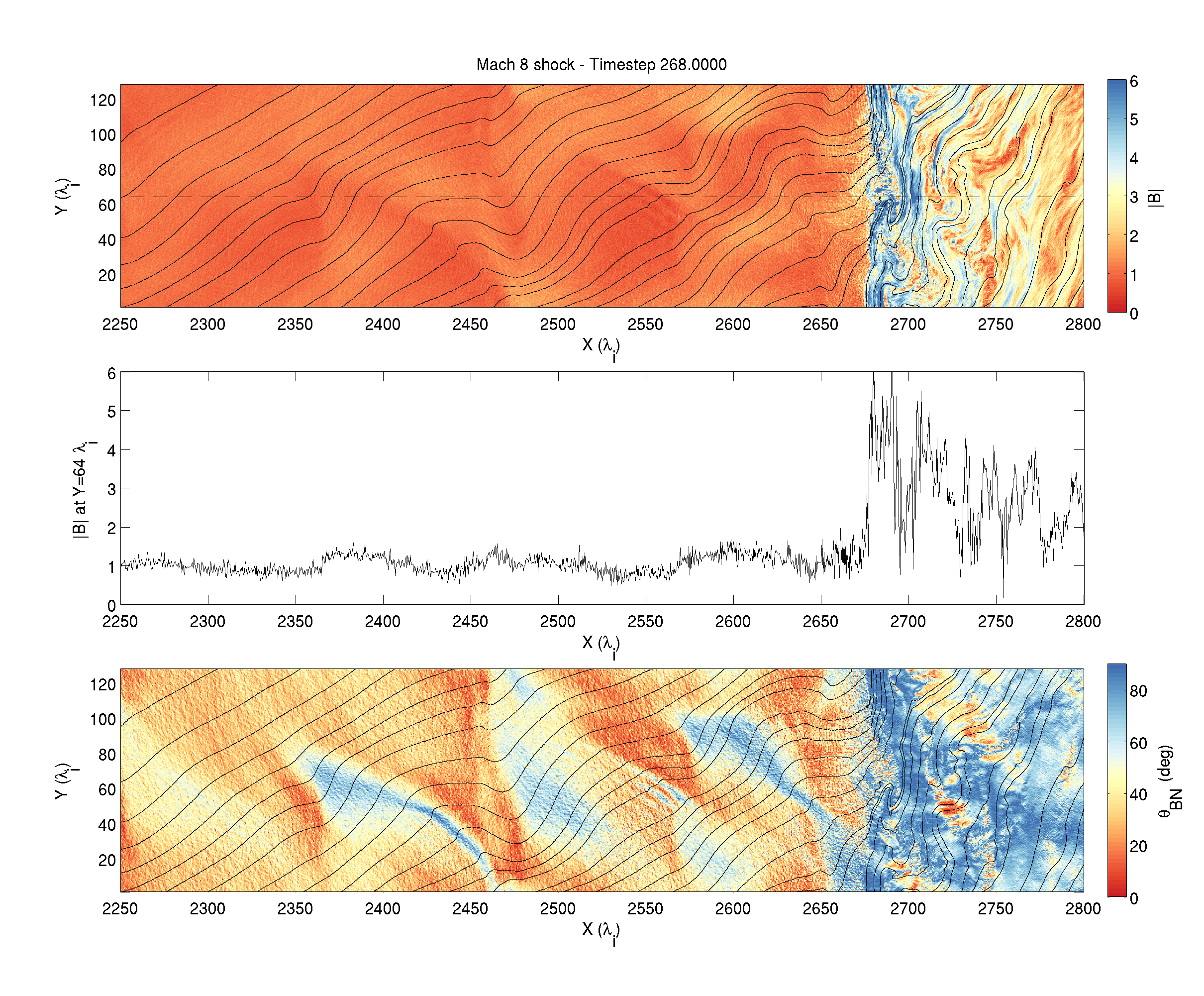}
\caption{Simulated magnetic field profile at time step 268 $\Omega_\mathrm{i}^{-1}$. The color scale in top panel shows the total magnetic field, together with the in-plane magnetic field lines (black). The second panel shows the magnetic field profile taken along $Y=64\lambda_\mathrm{i}$, indicated by the dashed line in the top panel. The third panel gives estimated values of the local $\theta_{Bn}$, calculated as the angle from the X-axis at each sampling point.}
\end{figure}

\begin{figure}
\includegraphics[angle=0,scale=.45]{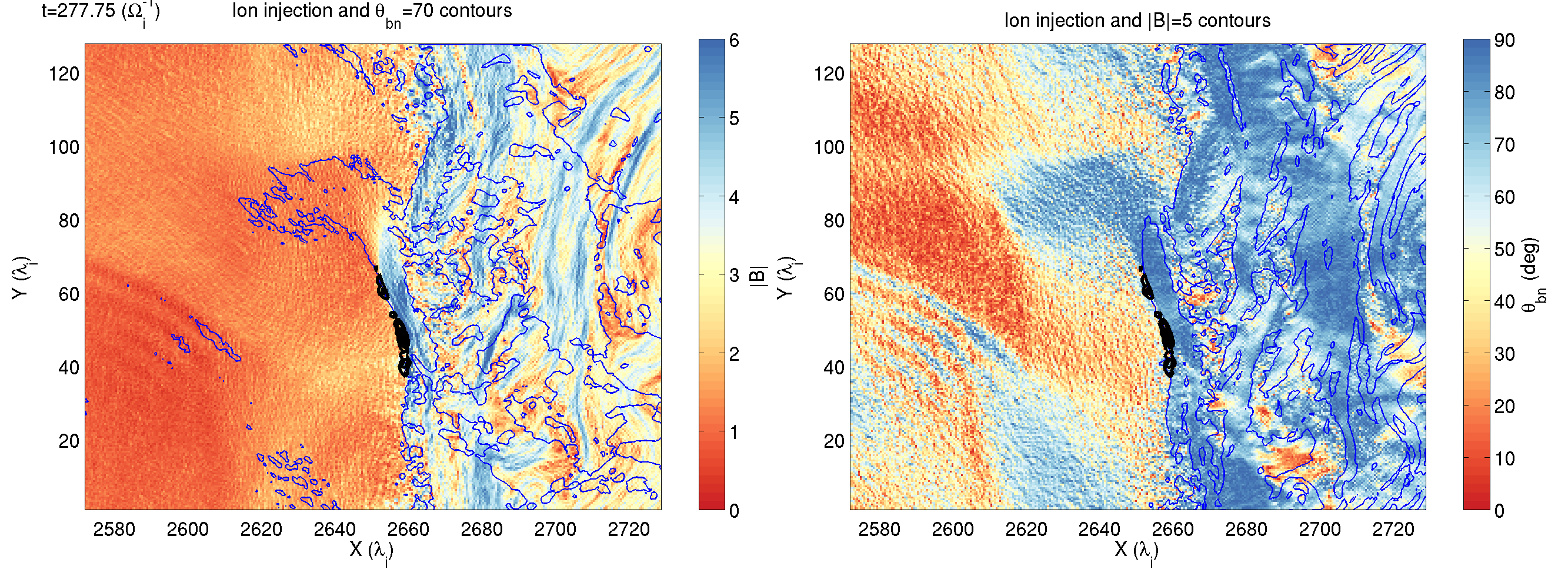}
\caption{Ion injections contours at $t$=277.75 $\Omega_\mathrm{i}^{-1}$. The left-hand panel shows the total magnetic field, with black lines showing  injection density contours, and blue lines showing regions of high $\theta_{Bn}$. In the right-hand panel, the color scales and the contours have been reversed, showing the local $\theta_{Bn}$ and regions of high magnetic fields, respectively.}
\end{figure}

\begin{figure}
\includegraphics[angle=0,scale=.52]{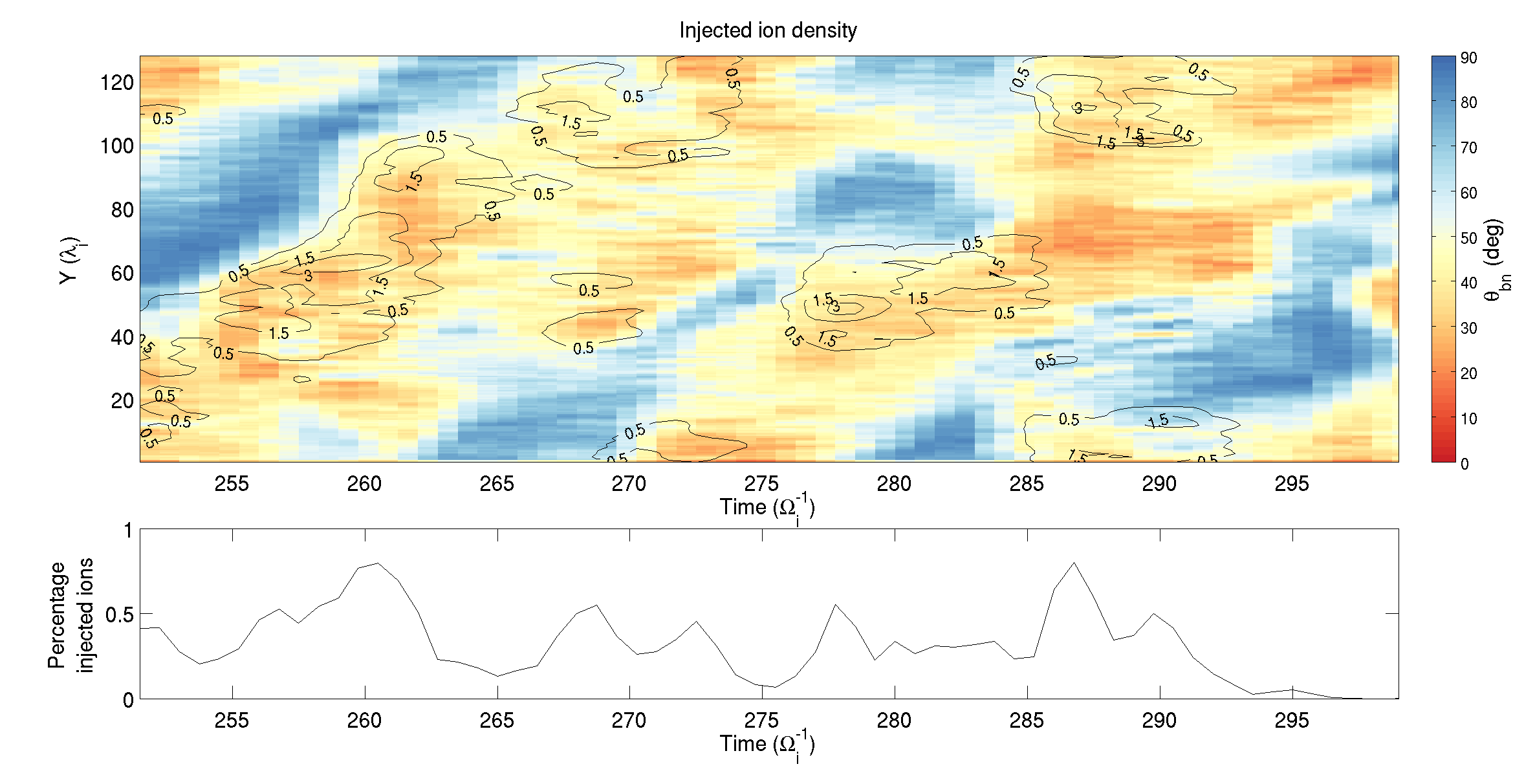}
\caption{Top panel: Ion injections contours are shown in black as a function of time, $Y$, and the upstream $\theta_{Bn}$ (color coded). Contour levels are given as the percentage compared to the average ion inflow. Bottom panel: The total percentage of injected ions at the shock as a function of time.}
\end{figure}

\clearpage

\begin{figure}
\includegraphics[angle=0,scale=.45]{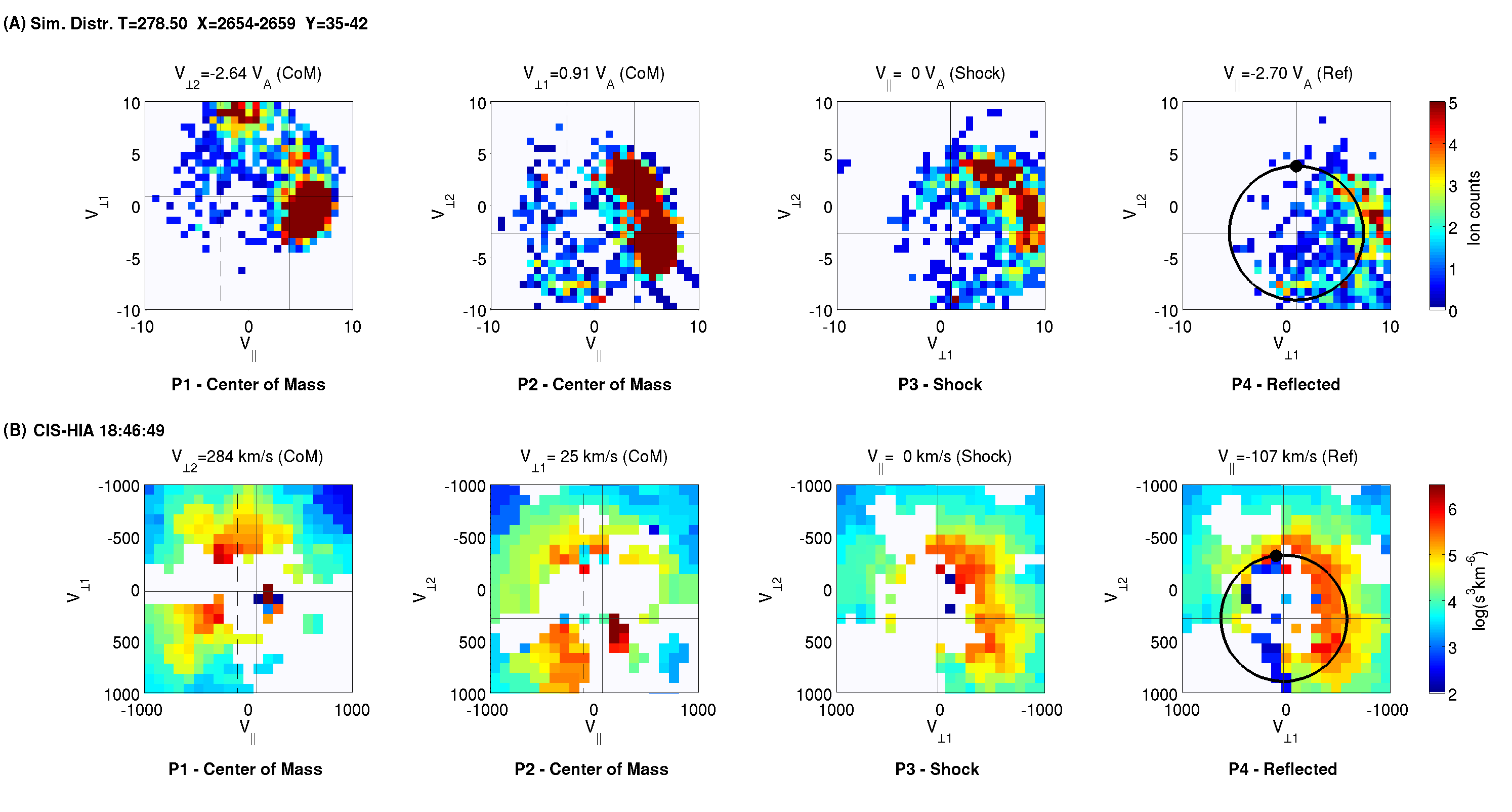}
\caption{Panel (A): Simulated velocity space distribution of a reflection event leading to ion injection. The figure follows the same format as Figure 3. An overview of the spatial and temporal evolution of this ion selection is given in Figures 8 and 9. Panel (B): The observed velocity distribution of a dispersive ion reflection event from Figure 3, rotated to match the coordinate system used in the simulations.}
\end{figure}

\begin{figure}
\includegraphics[angle=0,scale=.4]{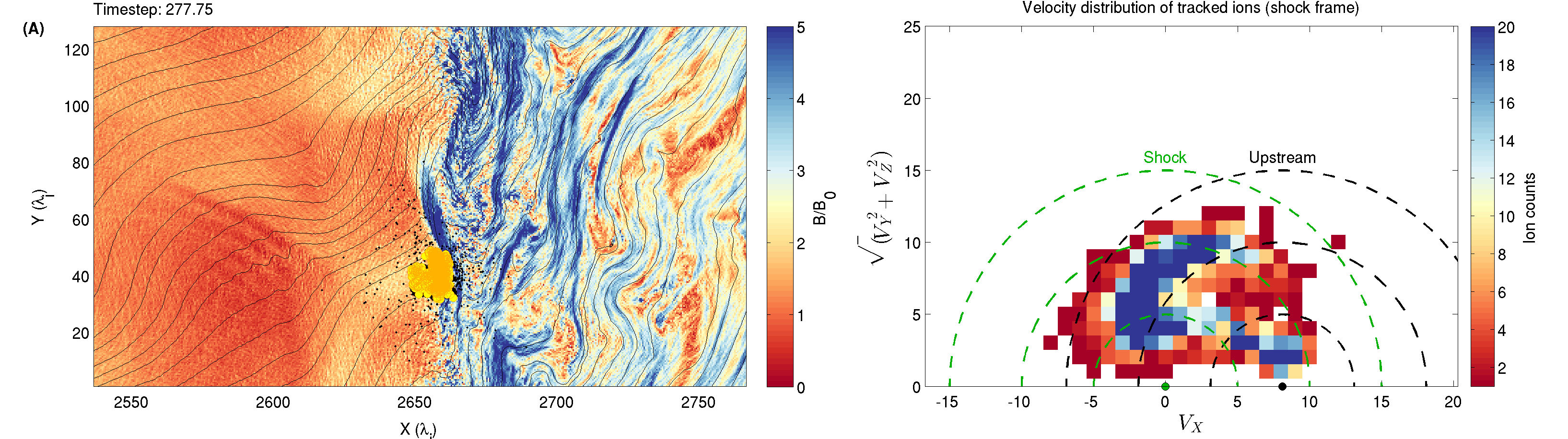}   
\includegraphics[angle=0,scale=.4]{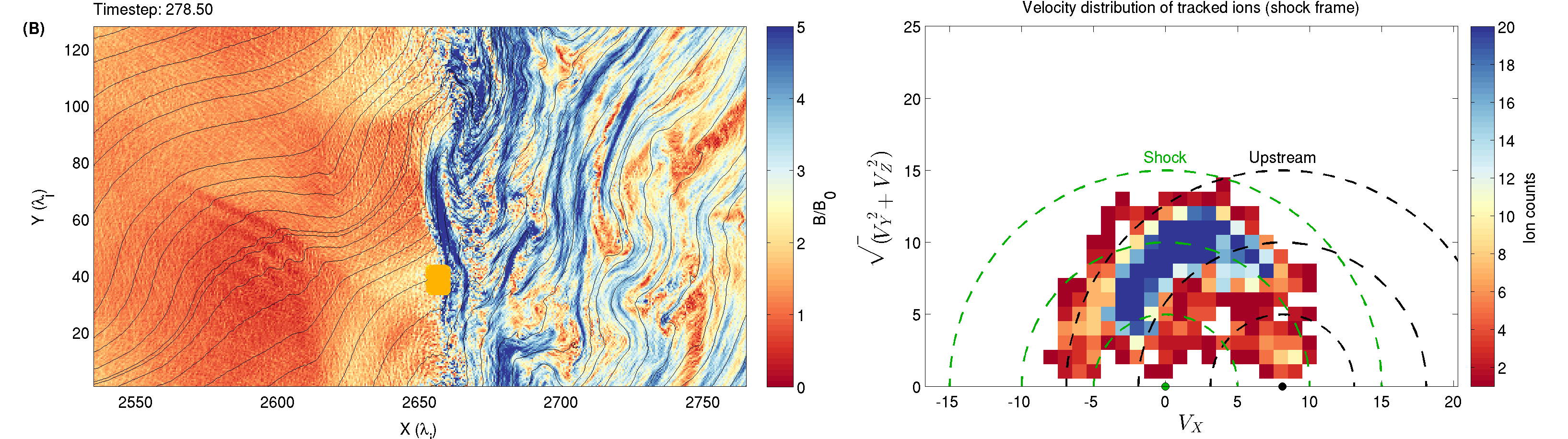}   
\includegraphics[angle=0,scale=.4]{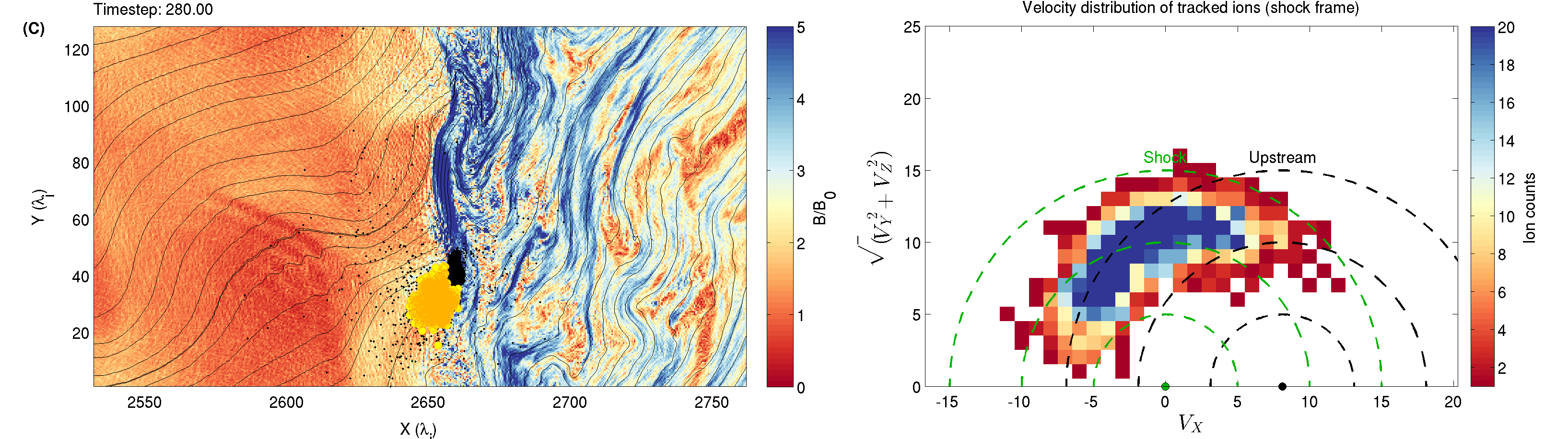}   
\caption{Three snapshots in time of the spatial (left column) and velocity space (right column) distribution of a selection of injected ions. The middle panel shows selection criteria in time and space. The yellow circles show the selection ion population, which are all found upstream of the shock near the end of the simulation. The black ion population in the background shows particles from the same spatial box in panel B, but that are either energized in the early stages of the simulation, or end up downstream of the shock. The velocity space distribution on the right is given for the injected (yellow) ion population only.}
\end{figure}

\clearpage

\begin{figure}
\includegraphics[angle=0,scale=.4]{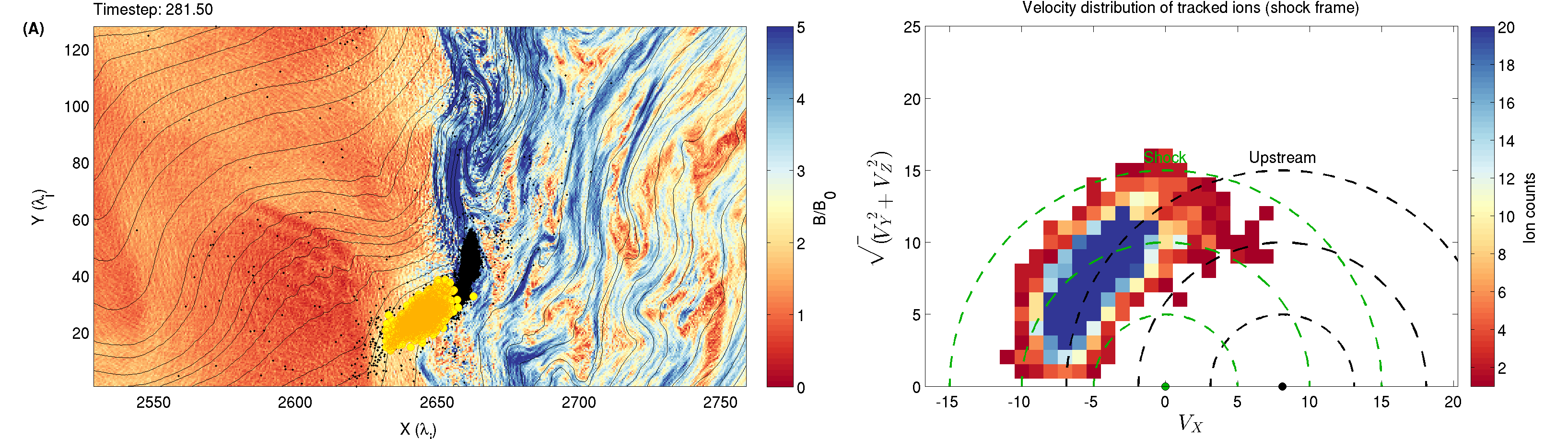}   
\includegraphics[angle=0,scale=.4]{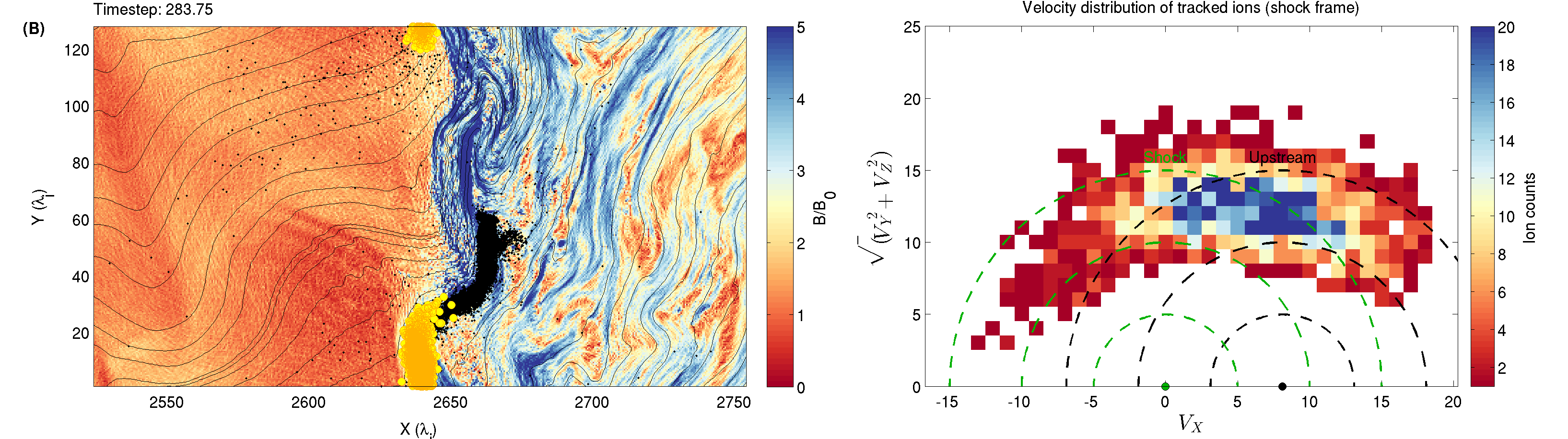}   
\includegraphics[angle=0,scale=.4]{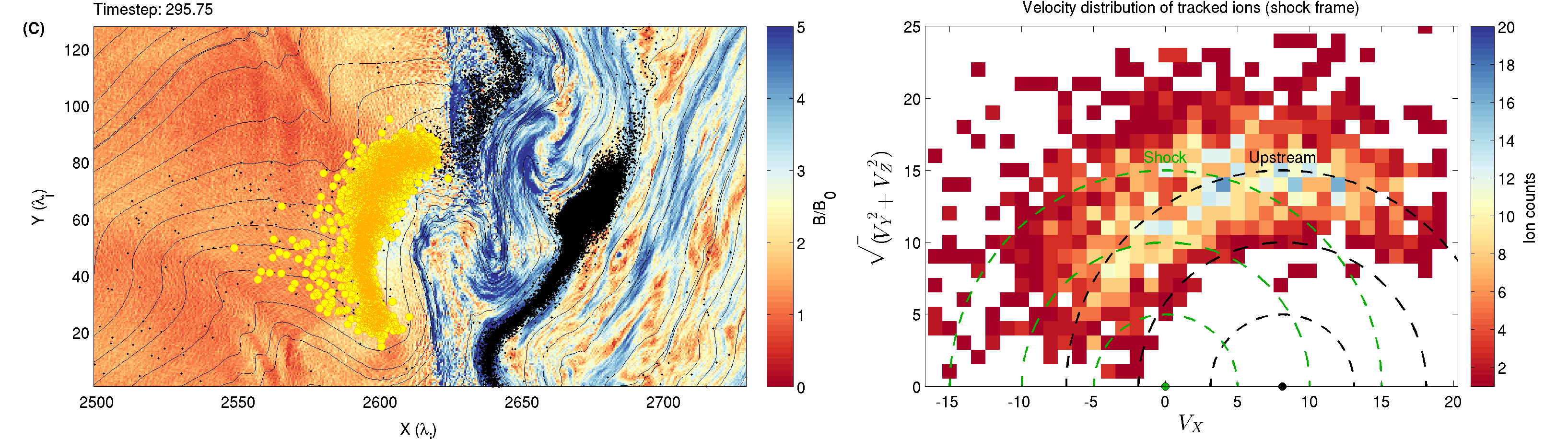}   
\caption{Three snapshots in time of the spatial (left column) and velocity space (right column) distribution of a selection of injected ions. The figure follows the same ion population as Figure 8, but at later time steps.}
\end{figure}

\clearpage

\begin{figure}
\includegraphics[angle=0,scale=.25]{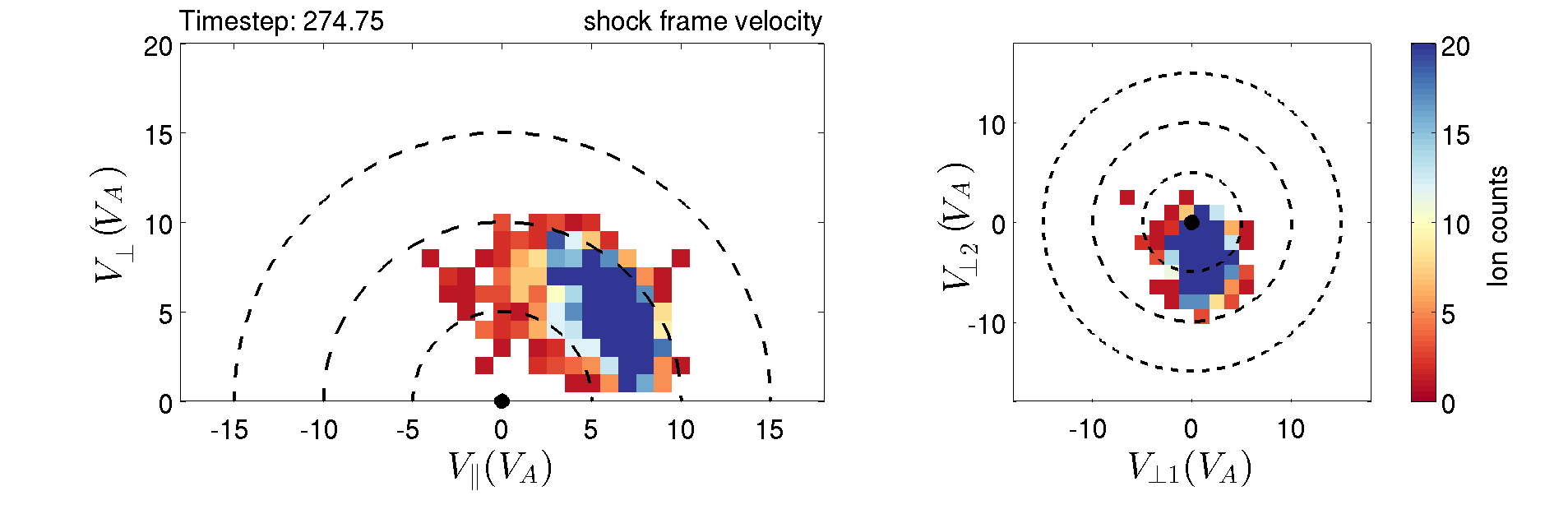}   
\includegraphics[angle=0,scale=.25]{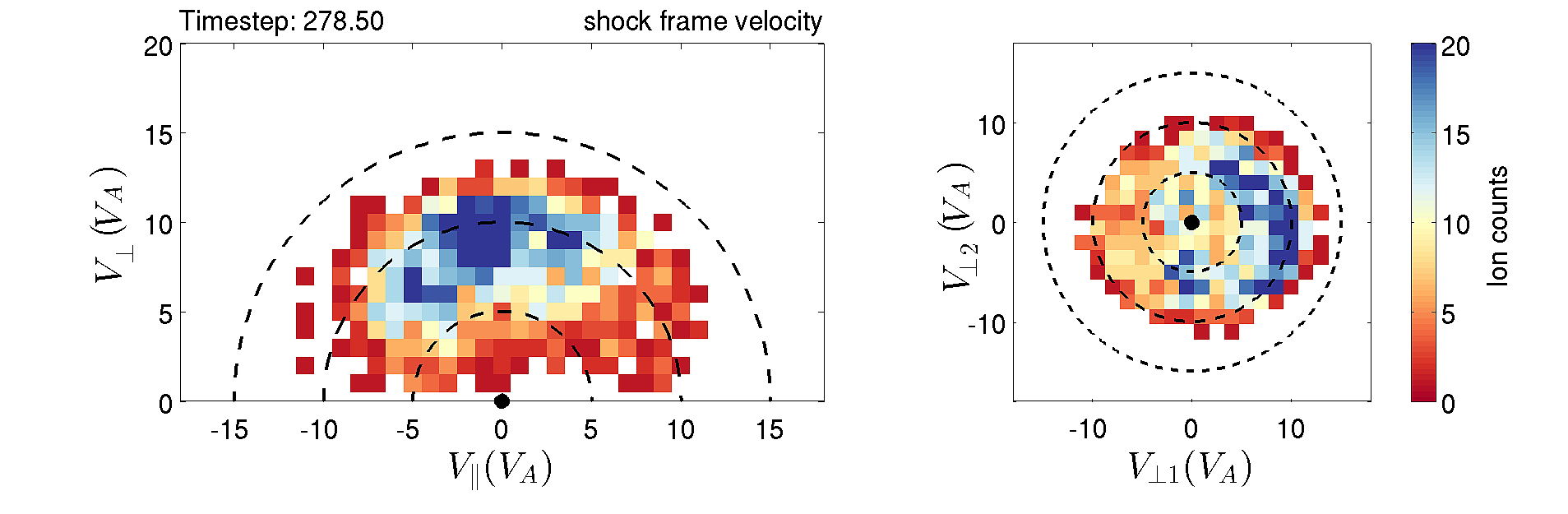}   
\includegraphics[angle=0,scale=.25]{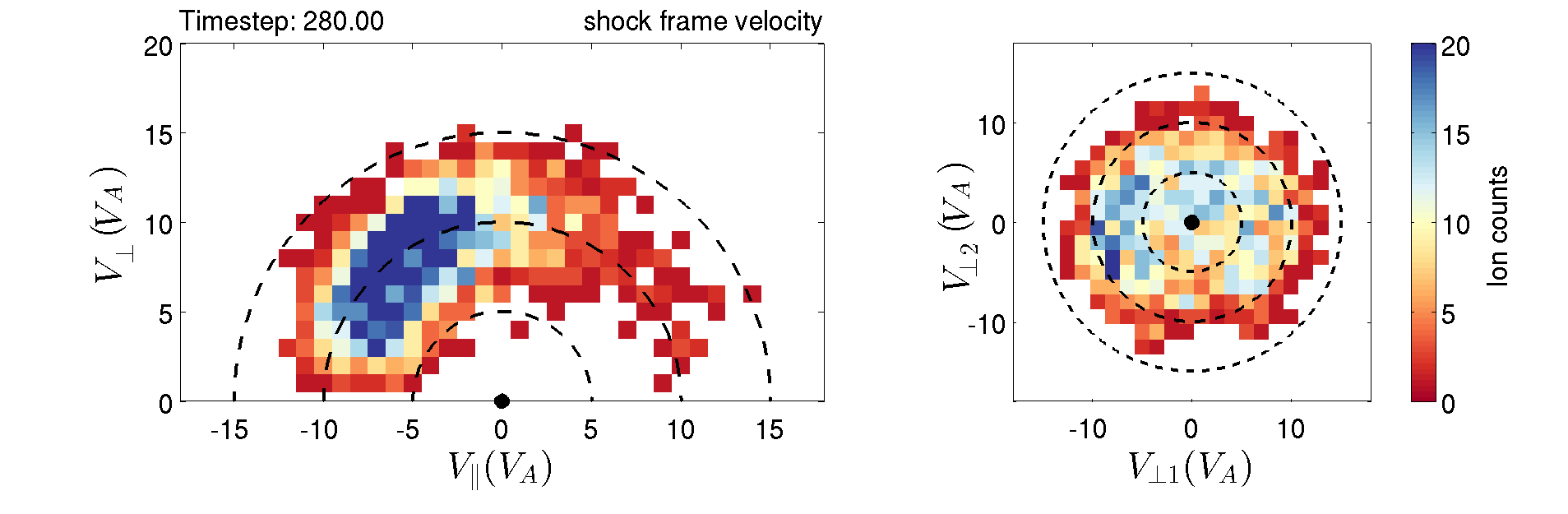}   

\caption{Evolution of the velocity space distribution in magnetic field-aligned coordinates of the injected ions shown in Figures 8 and 9. The figure shows how the injected ions change their field-aligned velocity from positive before the first interaction with the shock to negative after the shock interaction. This negative (upstream-directed) field-aligned velocity is essential for ion injection.}
\end{figure}

\end{document}